\documentclass{DJN}
\usepackage{graphicx}
\usepackage[utf8]{inputenc}
\usepackage[T1]{fontenc}
\usepackage{amsmath}
\usepackage{verbatim}
\usepackage{stmaryrd}
\usepackage{amsmath}
\usepackage{amsfonts}
\usepackage{bm}
\usepackage{amssymb}
\usepackage{graphicx}
\usepackage{latexsym}
\usepackage{amssymb}
\usepackage[utf8]{inputenc}
\usepackage[english]{babel}
\usepackage{mathrsfs}
\usepackage{caption}
\usepackage{subcaption}
\usepackage{breqn}
\usepackage{cancel}
\usepackage{calc}
\usepackage[title]{appendix}
\usepackage{rjwmath} 

\usepackage{float}
\usepackage{listings}
\usepackage[utf8]{inputenc}
\usepackage{xcolor}
\usepackage{lineno}
\definecolor{codegreen}{rgb}{0,0.6,0}
\definecolor{codegray}{rgb}{0.5,0.5,0.5}
\definecolor{codepurple}{rgb}{0.58,0,0.82}
\definecolor{backcolour}{rgb}{0.95,0.95,0.92}
\usepackage{todonotes}
\usepackage{tikz}
\usepackage{natbib}
\usepackage{autobreak}
\usepackage{esint}

\usepackage{array}

\newcolumntype{C}[1]{>{\centering\let\newline\\\arraybackslash\hspace{0pt}}m{#1}}

\newcolumntype{P}[1]{>{\centering\arraybackslash}p{#1}}

\usepackage[utf8]{inputenc}
\usepackage[T1]{fontenc}
\usepackage{amsmath}

\allowdisplaybreaks

\shorttitle{Modelling flow through a rapidly-oscillating elastic tube}
\shortauthor{D.J. Netherwood \& R.J. Whittaker}

\title{Modelling Fluid--Structure Interaction in an Initially Elliptical Elastic-Walled tube: Improved Onset Criterion for Self-Excited Oscillations}

\author{Daniel J. Netherwood{\aff{1,2}}
  \corresp{\email{daniel.netherwood@adelaide.edu.au}}
  \and Robert J. Whittaker\aff{1}}

\affiliation{\aff{1} School of Engineering, Mathematics and Physics, The University of East Anglia, Norwich Research Park, Norwich, United Kingdom, NR4 7TJ
\aff{2}School of Computer and Mathematical Sciences, University of Adelaide, Adelaide, SA 5000, Australia}

\begin{document}
\maketitle
\begin{abstract}
	We present a theoretical description of the fluid--structure interaction observed 
within a Starling resistor. The typical setup consists of a pre-stretched finite length 
thin-walled elastic tube mounted between two rigid tubes. The collapsible section is enclosed within a 
pressure chamber and a viscous fluid is driven through the system by imposing an axial 
volume flux at the downstream end. Valid within a long-wavelength thin-walled regime, we use our own results to 
model the wall mechanics.  These results arise from the solution of a generalised eigenvalue problem, 
and avoid the need to invoke the ad-hoc approximations made in previous studies. The wall 
mechanics are then coupled to the fluid mechanics using the Navier--Stokes equations, 
under the assumption that the oscillations in the tube wall are of small amplitude, 
long wavelength and high frequency.  We derive problems governing the leading-order steady and
oscillatory fluid-structure interaction.  At leading order, the
system permits normal-mode oscillations of constant frequency and
amplitude,  which are obtained in the form of series solutions.  Higher-order corrections govern the slow
growth or decay of the oscillations,  however (as in previous work)
these growth rates can be determined by analysing the system's global
energy budget without needing to compute the higher-order terms
explicitly.  Our results permit the first formal analysis of the errors incurred by neglecting contributions from higher-order azimuthal modes,  and enable the determination of improved criterion for the onset of self-excited oscillations in the tube wall.
	\end{abstract}
\section{Introduction}
Fluid flow through collapsible vessels can be observed throughout the biological and medicinal sciences.  Examples include the circulatory, respiratory,  lymphatic and central nervous systems.  In the circulatory system,  fluid flow through the vasculature facilitates the transportation of oxygen and nutrients to tissues and organs within the body \citep{pedley_1980}.   Additionally,  flow-induced deformations are a mechanism for the rupture of cholesterol deposits (plaques) inside of the arteries,  which can lead to potentially fatal vessel occlusion \citep{binns1989effect, ku1997blood}.  In the respiratory system,  the airways are deformable,  and forced expiration of air from the lungs can cause the airways to collapse \citep{macklem1971airway,  skalak1989biofluid}.  Considerable attention has been directed towards flow-induced instabilities arising from fluid--structure interaction in fluid-conveying elastic-walled tubes \citep{grotbertjensen,  heil2011fluid}.  Many physiological phenomena can be attributed to such instabilities.  Examples include: wheezing during forced expiration \citep{grotberg1989flutter,gavriely1989flutter},  Korotkoff sounds during sphygmomanometry \citep{bertram1989oscillations,  ur1970origin},  and cervical venous hum \citep{danahy1974cervical}.  

\begin{figure}
\begin{center}
\def\svgscale{0.3}
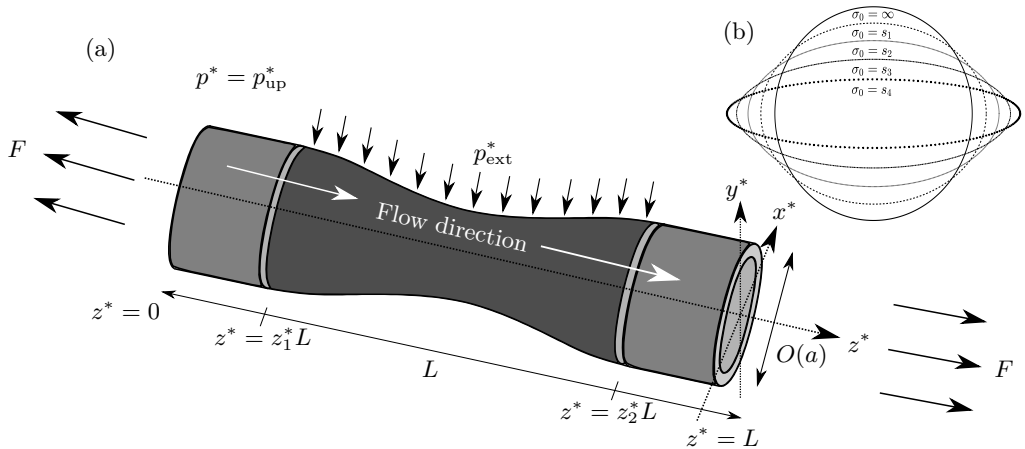
\end{center}
\caption{(a) The setup of an idealised Starling resistor. An initially elliptical elastic-walled tube is pinned between two rigid extensions. Fluid is driven through the system by imposing a steady dimensional axial volume flux of size $A_0^* \mathscr{U}$ at the downstream end $z^*=L$.  (b) The base-state ellipses corresponding to the representative ellipticity parameter values $\sigma_0 = \infty, s_1,s_2,s_3,s_4$.}\label{tube_picture}
\end{figure}

In this article,  we will focus on the theoretical modelling of flow through elastic-walled tubes.  Often,  these models are based on experimental investigations \cite[see,  e.g.  the review by][]{bertram2003experimental}.  Such experiments are typically performed within a `Starling resistor' \citep{knowlton1912influence}.  The setup consists of a thin-walled finite-length elastic tube. The tube is pre-stretched and clamped (at both ends) to two rigid tubes (see figure \ref{tube_picture}).  Fluid is driven through the system either by imposing a pressure difference between the ends of the tube, or by imposing a flow rate at one end through a volumetric pump.  By enclosing the collapsible section of the tube inside of a pressure chamber,  an external pressure,  $p^*_{\mathrm{ext}}$,  can be applied to the tube's outer surface \citep{bertram1986unstable, bertram1990mapping}.  Deformations in the tube wall then occur due to the combined effect of the fluid traction (i.e.,  the internal hydrodynamic pressure $p^*_{\mathrm{int}}$ and viscous shear forces) and the applied external pressure $p^*_{\mathrm{ext}}$.  For large-Reynolds-number flows,  contributions from the viscous shear stresses are dominated by inertial effects,  and are often neglected.  In this regime,  deformations are then said to take place due to the  transmural pressure $p^*_{\mathrm{tm}}= p^*_{\mathrm{int}}-p^*_{\mathrm{ext}}$,  which is the pressure difference between the inside and outside of the tube.  Experiments reveal that for sufficiently negative transmural pressures,  the tube will buckle into an elliptical-like configuration.  In this deformed state,  small changes in the transmural pressure yield large changes in the cross-sectional area of the tube.  Provided that the mean axial flow rate is large enough that the net influx of kinetic energy into the system is sufficiently large as to overcome viscous losses,  the system can exhibit high and low frequency self-excited oscillations (of large and small amplitude) associated with a number of different instability mechanisms.  

For the case of two-dimensional channel flow,  developments have been made in producing theoretical and numerical models of the Starling resistor that predict self-excited oscillations. The first model that gathered considerable traction was that of \citet{pedley1992longitudinal}, who formulated the problem of fluid flow (driven by an imposed pressure drop) through a two-dimensional planar channel in which one wall has a section replaced by an elastic membrane held under longitudinal tension.  Provided that the mean-flow Reynolds number is large,  it has been shown that this system exhibits a rich variety of flow-induced instabilities  \citep[see numerical studies by][]{rast1994simultaneous,  luo1995numerical,  luo1996numerical,  luo2000multiple}.   

In the large-membrane-tension regime,  the steady viscous pressure drop induces small-amplitude wall deflections,  and the system is susceptible to small-amplitude high-frequency self-excited oscillations. \citet{jensen2003high} were the first to produce a theoretical description of these oscillations.  For a regime in which the membrane tension, mean-flow Reynolds number and axial lengthscales were all large, \citet{jensen2003high} formally identified a `sloshing' mechanism that drives the self-excited oscillations, and deduced a stability threshold which can be used to predict their onset.  It is this mechanism that we focus on in this article.

We now give a brief description of how the sloshing mechanism identified by \citet{jensen2003high} can result in the onset of self-excited oscillations. Owing to the large longitudinal tension within the membrane, the deflections in the compliant wall are small, and large elastic restoring forces are exerted on the internal fluid. The small-amplitude displacements in the flexible wall change the volume of the channel and displace fluid particles (periodically) towards the upstream and downstream ends of the tube, resulting in oscillatory `axial sloshing' flows in the rigid sections.  By virtue of a non-zero time-mean-square,  any oscillatory flow at the upstream end increases the kinetic energy influx and any oscillatory flow at the downstream end increases the kinetic energy outflux.  Hence,  if the amplitude of the oscillations is greater in the upstream section of the tube, then there will be a net influx of kinetic energy into the system.  Provided that this input exceeds additional losses (e.g., the dissipation due to the viscosity in the fluid and work done by the pressure at the tube ends),  then this additional kinetic energy can be sufficient to drive  the instability \citep{jensen2003high, heil2008rapidly}.  Alternative instability mechanisms also exist,  which do not necessarily rely on an increase in kinetic energy flux,  but instead a minimisation of viscous losses or a reduction in the work done by the pressure at the tube ends \citep{stewart2009local}.

\citet{whittaker2010predicting} were the first to construct a three-dimensional theoretical model of the high-frequency self-excited oscillations observed in a Starling resistor. They investigated the problem of an elastic-walled tube with an initially axially uniform elliptical cross section conveying an incompressible viscous fluid. To derive the model,  \citet{whittaker2010predicting} combined their own asymptotic descriptions for the fluid mechanics \citep{whittaker_waters_jensen_boyle_heil_2010} and wall mechanics \citep{rationalderivationofatubelaw}.  

\citet{whittaker_waters_jensen_boyle_heil_2010} showed that the conservation of mass and axial momentum leads to a system of two differential equations relating the dimensionless axial fluid velocity,  $w$, fluid pressure $p$,  and cross-sectional area of the tube $A$.   Using shell theory, \citet{rationalderivationofatubelaw} derived an explicit relationship (known as a tube law) between  $A$ and $p-p_{\mathrm{ext}}$,  where $p_{\mathrm{ext}}$ is the (known) steady dimensionless external pressure.  Overall,  this means that their asymptotic model of the fully coupled problem involves a system of three equations relating three dependent variables $w,p$ and $A$,  which depend only on the dimensionless axial coordinate $z$ and time $t$.  \citet{whittaker2010predicting} manipulated this system to eliminate $w$ and $A$ in favour of the pressure,  $p$.  This problem was solved to find the leading-order normal modes of the system.  The system's energy budget was then used to compute the slow growth rates of these modes,  demonstrating that the system can exhibit self-excited oscillations.  They compared their results with direct numerical simulations and obtained good agreement. The model of \citet{whittaker2010predicting} was then extended by \citet{walters2018effect} to include effects due to the inertia of the tube wall. It was found that the addition of wall inertia has a stabilising effect on the system. 

Whilst the work of both \citet{whittaker2010predicting} and \citet{walters2018effect} are significant improvements on previous attempts to model the Starling resistor, the method used to derive the tube law --- which couples the fluid and wall mechanics --- has its limitations. By projecting the azimuthal displacement onto a basis of azimuthal Fourier modes, an adhoc approximation based on the relative sizes of each azimuthal mode was used (essentially truncating at $n=1$), allowing for the derivation of the tube law. The main problem with this approach is the difficulty in calculating the relative error after neglecting contributions from higher modes, due to the inherent coupling between the modes.  \citet{Netherwood2023deformations} derived a system of generalised tube laws that includes contributions from the higher-order azimuthal modes by instead projecting the solution onto a basis of eigenfunctions, resulting from a generalised eigenvalue problem. This new method resulted in a formal series solution for the solid-mechanical problem,  whilst significantly simplifying the calculations required to compute the relative error incurred by truncating the solution after any azimuthal mode.  

In this article,  we consider the fluid--structure interaction problem of a viscous fluid conveyed within a long thin-walled elastic tube which has an initially axially uniform elliptical cross section.  We model the wall mechanics using the improved solid-mechanical model recently derived by \citet{Netherwood2023deformations},  and then adapt the methodology of \citet{whittaker2010predicting} and \citet{walters2018effect} to couple this to the fluid mechanics.  We find that the retention of higher-order azimuthal modes in the analysis results in a more complicated system of governing equations for the fluid--structure interaction,  owing to the fact that the azimuthal modes do not decouple at leading order.  Area displacements that are associated with the first azimuthal eigenmode create a pressure distribution in the fluid that forces all of the azimuthal eigenmodes. We show that we can overcome this limitation by observing that the fundamental azimuthal mode dominates the area displacements. This means that the dominant contribution to the pressure in the fluid is forced by the $n=1$ azimuthal mode, and that the response from this pressure is to excite predominantly the first azimuthal mode.  This analysis results in weak coupling between the higher-order azimuthal modes. We show that this simplification enables us to adopt a series expansion for the normal modes and oscillation frequencies of the system.  

We organise this article as follows. In \S\ref{set_up_section_coupled},  we provide a full description of the physical setup. The problem is decomposed into steady and oscillatory parts and then non-dimensionalised. The parameter regime in which our model is considered is then presented. In \S \ref{modelling_section},  we introduce the models used to describe the fluid and wall mechanics respectively.  In \S \ref{unsteady_coupled_problem_section},  we derive governing equations and boundary conditions for the oscillatory normal modes that describe the leading-order fluid--structure interaction in the tube.  Solutions of these equations are then sought in the form of series expansions.  In \S \ref{stability_section},  we use the global energy budget to compute the growth rate of the normal modes,  and hence determine an improved stability threshold for the onset of self-excited oscillations in the tube wall.  This is expressed as a critical mean-flow Reynolds number above which an instability will grow.  Finally, in \S \ref{Conclusion_section}, we discuss our results and comment on potential future work. 
\section{Setup}\label{set_up_section_coupled}
\subsection{Problem description}\label{problem_description_section}
We adopt the setup of \cite{Netherwood2023deformations} by considering a thin-walled tube of dimensional length $L$ and circumference $2\pi a$ (see figure \ref{tube_picture}). The tube has an initially axially uniform elliptical cross section, which is aligned with dimensional Cartesian co-ordinates $(x^*,y^*,z^*)$ such that the tube's centreline lies along the $z^*$ axis.  We also introduce $t^*$ as dimensional time. The major and minor axes of the tube's cross section are aligned with the $x^*$ and $y^*$ axes respectively. The ellipticity of the tube's cross section is set by the parameter $\sigma_0$ such that the major and minor axis of the cross section are $ac \cosh \sigma_0$ and $ac \sinh \sigma_0$ respectively, where 
\begin{equation}
c(\sigma_0)= \frac{\pi \sech \sigma_0}{2 E(\sech \sigma_0)}
\end{equation}
is a normalisation factor, which is introduced to set the tube's initial circumference to be $2\pi a$. Here 
\begin{equation}
E(k)= \int_0^{\pi/2}(1-k^2 \sin^2 \phi)^{1/2}\D{\phi}
\end{equation}
is the complete elliptic integral of the second kind. The tube's initial dimensional cross-sectional area,  $A_0^*$,  can then be calculated as
\begin{equation}
A_0^*=\pi a^2c^2 \sinh \sigma_0 \cosh \sigma_0= \pi a^2 \frac{\pi^2 \tanh \sigma_0}{4[E(\sech \sigma_0)]^2}.
\end{equation}
Throughout this paper we shall refer to a set of four representative values of $\sigma_0$,  which were introduced by \cite{Netherwood2023deformations}.  They are given by $\sigma_0 = s_i$, where $s_1 = 0.9540, s_2 = 0.6, s_3 = 0.3840, s_4 = 0.2194$. The corresponding elliptical cross sections are shown in figure \ref{tube_picture}.

In accordance with the experimental setup of the Starling resistor, the tube comprises an elastic section of material having dimensional mass per unit area,  $m$,  and wall thickness,  $d$,  occupying $z_1 L<z^*<z_2L$, which is pinned between two rigid sections occupying $0<z^*<z_1L$ and $z_2L<z^*<L$.  In the elastic section, the tube is able to deform in response to the combined effect of the steady dimensional external pressure $p_{\mathrm{ext}}^*$ and the fluid traction. Since the elastic section of the tube is pinned between two rigid sections, a dimensional axial tension force, $F$, can be imposed at the two ends of the tube.  This results in a uniform axial pre-stress of magnitude $F/(2 \pi a d)$.  We consider deformations in the tube wall about the elliptical axially uniform pre-stressed state. We assume that the elastic section of the tube is linearly elastic and behaves isotropically with incremental Young's modulus $E$ and Poisson ratio $\nu$. The bending stiffness is defined as \begin{equation}
K=\frac{Ed^3}{12(1-\nu^2)}.
\end{equation}

We investigate the case in which the tube conveys an incompressible viscous fluid of density $\rho$ and dynamic viscosity $\mu$. The fluid is driven through the system by imposing a steady dimensional axial volume flux of size $A_0^*\mathscr{U}$ at the downstream end.  At the upstream end, we fix the dimensional pressure $p^*=p_{\mathrm{up}}^*$.  These boundary conditions are chosen so that the amplitude of the resulting oscillatory axial sloshing flow at the downstream end is zero. This ensures no kinetic energy is lost there,  which increases the likelihood that an instability will occur.  We denote the dimensional axial fluid velocity component as $w^*$ and the dimensional transverse fluid velocity vector as $\bm{u}_\perp^*$.  We define $A^*$ as the dimensional cross-sectional area of the tube in its deformed configuration.  

\subsection{Oscillatory time scale}
We consider the case of oscillations in the tube wall of slowly varying amplitude having typical (normal) amplitude $b(t^*) \ll a$, and oscillation time scale $T$.  Hence,  the scale for the normal velocity of the wall is $b/T$.  It is then natural to take $b/T$ as the scale for the transverse oscillatory velocity in the fluid.  The scale for the axial oscillatory velocity of the fluid is then estimated as $bL/(aT)$ through continuity. 

We can formulate an explicit expression for the time scale $T$ by assuming that the oscillations arise due to a balance between axial fluid inertia and restoring forces from azimuthal bending of the tube wall.\footnote{In the regime specified in \S\ref{parameter_regimes},  there are two principle restoring forces to deformations of the tube wall: azimuthal bending and axial tension-curvature.  By the assumption $\tilde{F}=O(1)$ in (\ref{eq: solid_mechanics_regime}), their magnitudes are comparable,  so either could be used here to estimate the timescale. } The pressure scale associated with the forces due to azimuthal bending  is $Kb/a^4$. Equating this with the unsteady axial inertial pressure scale $\rho L^2 b/(aT^2)$, we find that
\begin{equation}
T = \bigg( \frac{\rho a^3 L^2}{K} \bigg)^{1/2}.
\end{equation}
\subsection{Dimensionless groups and parameter regimes}\label{parameter_regimes}
The setup described above gives rise to seven independent dimensionless groups. There are three quantities associated with the geometry of the tube, which correspond to wall thickness, tube length and oscillation amplitude:
\begin{equation}
\delta = \frac{d}{a} \qquad \ell = \frac{L}{a}, \qquad \Delta=\frac{b(t^*)}{a}.
\end{equation}
There are two independent dimensionless groups associated with the fluid mechanics. We follow  \citet{whittaker2010predicting} and define the Womersley number $\alpha$ and the Strouhal number $St$ as:
\begin{equation}
{\alpha}^2 \equiv \frac{\rho a^2}{\mu T}=\bigg(\frac{\rho K}{a \ell^2 \mu^2}\bigg)^{1/2} \qquad \text{and} \qquad St \equiv \frac{a}{\mathscr{U} T}=\bigg(\frac{K}{\rho a^3 \mathscr{U}^2}\bigg)^{1/2}.
\end{equation}
The Womersley number $\alpha$ measures the relative importance of unsteady inertia to viscous effects. The Strouhal number $St$ measures the relative importance of unsteady to convective inertia. The steady Reynolds number, with its usual definition,  can be expressed in terms of $\alpha$ and $St$ by
\begin{equation}
Re = \frac{\rho \mathscr{U} a}{\mu}= \frac{{\alpha}^2}{St}.
\end{equation}
There are two dimensionless groups associated with the solid mechanics of the tube wall. We define the dimensionless axial tension $\tilde{F}$ and inertia coefficient (or dimensionless wall mass) $M$ as follows:
\begin{equation}
\tilde{F}=\frac{aF}{2 \pi K \ell^2}, \qquad M=\frac{ma^4}{KT^2}\equiv \frac{m}{\rho a \ell^2}.
\end{equation}
The dimensionless axial tension measures the ratio of the restoring forces due to axial curvature/tension effects $Fb/(2\pi a L^2)$ and azimuthal bending $(Kb/a^4)$ respectively. The inertia coefficient was introduced by \citet{walters2018effect} and is defined as the ratio between forces due to wall inertia $(mb/T^2)$ and azimuthal bending or equivalently, the forces due to axial fluid inertia $(\rho a \ell^2 b/T^2)$. 

Following the formulation of \citet{whittaker2010predicting}, we work within a parameter regime in which the tube is long and thin, subject to a large axial tension such that the tube exhibits small amplitude, high-frequency deformations of long wavelength. Mathematically, these assumptions correspond to:
\begin{equation}\label{parameter_regimes_eq}
\ell \gg 1, \qquad \delta \ll 1, \qquad \Delta\ll 1, \qquad \alpha \gg 1, \qquad \ell St \gg 1.
\end{equation}
We also adopt a regime in which tension/curvature effects are of equivalent magnitude to azimuthal bending effects, enabling both to feature at leading order.  We also permit inertial effects from the tube wall,  which may at most balance the restoring force from azimuthal bending.  Together,  these correspond to: 
\begin{equation}\label{eq: solid_mechanics_regime}
\tilde{F}=O(1) \qquad \text{and} \qquad M \lesssim 1. 
\end{equation}
We assume that the time scale for the growth/decay of the oscillations is sufficiently large so that (slow) changes to the non-oscillatory fluid flow and wall deformation do not enter the problem at leading order.  For the flux conditions considered here,  this will always be the case when the flow rate is close to the neutrally stable critical value.  For a detailed discussion of this topic,  see \citet{whittaker2011energetics}.  
\subsection{Non-dimensionalisation and scaling}\label{scale_analysis_section}
We follow the non-dimensionalisation of \citet{whittaker2010predicting}. Axial lengths are scaled with the tube length $L$, transverse lengths with the radial scale $a$, and time with the estimated timescale $T$. We write
\begin{equation}\label{co-ordinate_scales}
(x^*,y^*,z^*)= (ax,ay,Lz), \;\;\; A_0^* = a^2 A_0, \;\;\; A^*= a^2 A, \;\;\; t^*=Tt,
\end{equation}
where the unstarred variables are the non-dimensional counterparts of the starred variants. 

The velocity and pressure of the fluid are decomposed into their respective steady and oscillatory components. For the steady component, the axial scale is the mean flow $\mathscr{U}$, and the transverse scale $\mathscr{U}a/L$ arises from continuity and the tube's aspect ratio. For the oscillatory component, the transverse scale is the normal velocity scale $b/T$ of the wall motion, and the axial scale $bL/aT$ arises from continuity. When non-dimensionalising the pressure, the viscous scale is used for the steady component, and the oscillatory component is scaled with unsteady axial inertia. We therefore write:
\begin{align}
\bm{u}_\perp^*& = \frac{\mathscr{U}a}{L}\bar{\bm{u}}_\perp +\frac{b}{T} \hat{\bm{u}}_\perp= \frac{\mathscr{U}}{\ell}\left(\bar{\bm{u}}_\perp +\ell St \Delta \hat{\bm{u}}_\perp \right), \label{transverse_velocity_scales}\\ 
w^*&= \mathscr{U}\bar{w}+\frac{Lb}{aT}\hat{w}=\mathscr{U}\left(\bar{w}+\ell St\Delta \hat{w} \right), \label{axial_velocity_decomposition}\\  p^*-p_{\mathrm{up}}^*&= \frac{\mu L \mathscr{U}}{a^2}\bar{p}+\frac{\rho L^2 b}{aT^2}\hat{p}=\frac{\mu L \mathscr{U}}{a^2}\left(\bar{p}+\alpha^2 \ell St \Delta \hat{p}\right), \label{pressure_decomposition} 
\end{align}
where overbars denote the steady components and hats the unsteady components.
The dimensional external pressure $p_{\mathrm{ext}}^*$ (assumed to be steady) is non-dimensionalised on the steady viscous scale as 
\begin{equation}
p_{\mathrm{ext}}^*-p_{\mathrm{up}}^*=\frac{\mu L \mathscr{U}}{a^2}\bar{p}_{\mathrm{ext}}.
\end{equation}
Using this expression, together with (\ref{pressure_decomposition}), we find that the dimensional transmural pressure $p_{\mathrm{tm}}^*$ can be written as
\begin{align}
p_{\mathrm{tm}}^*= p^*-p_{\mathrm{ext}}^*= \frac{K}{a^3}\left(\frac{1}{\alpha^2 \ell St} \left(\bar{p}-\bar{p}_{\mathrm{ext}}\right)+\Delta(t) \hat{p}\right) = \frac{\Delta K}{a^3} p_{\mathrm{tm}},\label{dimensional_transmural}
\end{align}
where 
\begin{equation}
p_{\mathrm{tm}} = \frac{1}{\Delta \alpha^2 \ell St}\left(\bar{p} - \bar{p}_{\mathrm{ext}} \right) +\hat{p},
\end{equation}
and $\bar{p}$ and $\hat{p}$ are evaluated at the tube wall.

Energy and energy fluxes are non-dimensionalised respectively on
\begin{equation}
\rho \mathcal{U}^2 a^2 L \qquad \text{and} \qquad \rho \mathcal{U}^3a^3.
\end{equation}

The tube wall deforms in response to the transmural pressure, which in turn leads to changes in the cross-sectional area. The area is non-dimensionalised as 
\begin{align}\label{area_expression}
A^*(z,t)&= a^2 A(z,t) = a^2\left(A_0+\frac{1}{{\alpha}^2 \ell St}\bar{A}(z)+\Delta(t) \hat{A}(z,t)\right),
\end{align}
in terms of steady and oscillatory perturbations. The oscillatory perturbation has been scaled using the natural scale $(ab = a^2\Delta)$ from wall motion, whilst the steady perturbation is set to ensure that the ratio between the scales for steady and unsteady perturbations is the same as for the transmural pressure in (\ref{dimensional_transmural}).
\section{Mathematical modelling}\label{modelling_section}
We now present a description of the leading-order models used to describe the fluid and solid mechanics present in the problem described above.  For the fluid mechanics,  we adopt the modelling of \cite{whittaker2010predicting},  who produced an asymptotic description of flow within the tube based on the Navier--Stokes equations by exploiting the small amplitude,  high frequency and long-wavelength nature of the oscillations. For the wall mechanics, we adopt the solid mechanical model of \cite{Netherwood2023deformations},  who derived an expression for the relative change of the tube's cross-sectional area $A^*-A_0^*$ in terms of the transmural pressure $p_{\mathrm{tm}}^*$. The fluid and solid mechanics are then coupled via $A^*$ and $p_{\mathrm{tm}}^*$.
\subsection{Fluid mechanics}\label{fluid_mechanics_section}
To model the fluid inside of the tube,  we follow \cite{whittaker2010predicting} and \cite{walters2018effect} who,  for the parameter regimes considered here,  determined governing equations for the steady and oscillatory axial fluid velocity,  fluid pressure and cross-sectional area.  The calculations are the same as those presented in those works,  so we omit the details for brevity.  
\subsubsection{Equations for the steady flow}
For the steady component, it is found that $\bar{p}$ is uniform within each cross section at leading order. The steady component of the axial momentum equation then yields the following leading-order balance between the axial pressure gradient and viscosity
\begin{equation}\label{steady-axial_fluid}
\nabla^2\bar{w}=\frac{\D{\bar{p}}}{\D{z}}.
\end{equation}
(Nonlinear inertia and contributions from the oscillatory flow only enter at higher orders.)

We also have the leading-order steady component of the cross-sectionally integrated continuity equation
\begin{equation}\label{steady_continuity_equation}
\frac{\D{}}{\D{z}}\iint_{\bar{\mathscr{A}}(z)} \bar{w}\D{S}=0,
\end{equation}
where $\bar{\mathscr{A}}$ is the space enclosed by the dimensionless mean position of the wall.  

At the upstream end ($z=0$), the fluid pressure is fixed as $p^*=p^*_{\mathrm{up}}$.  Using (\ref{pressure_decomposition}), the steady component of this condition is
\begin{equation}\label{upstreamb.c}
\bar{p}=0 \qquad \text{at} \qquad z = 0.
\end{equation}

At the downstream end ($z=1$), the axial volume flux is fixed. Using (\ref{axial_velocity_decomposition}),  the steady contribution to this boundary condition is given by
\begin{equation}\label{downstream_b.c}
\iint_{\bar{\mathscr{A}}(z)} \bar{w} \D{S} = A_0 \qquad \text{at} \qquad z=1.
\end{equation}
We also have the no-slip boundary condition
\begin{equation}\label{tube_wall_b.c1}
\bar{w}=0 \qquad \text{on the tube wall}.
\end{equation}
\subsubsection{Equations for the oscillatory flow}
For the oscillatory component, the long wavelength and high-frequency nature of the oscillations ($\ell, \alpha \gg 1)$ result in both the axial velocity and pressure being cross-sectionally uniform (outside of viscous boundary layers) at leading order.  The axial component of the momentum equation reduces to an inertial balance between axial fluid velocity and axial pressure gradient
\begin{equation}\label{oscialltory_inertial_balance}
\frac{\partial\hat{w}}{\partial t}=-\frac{\partial \hat{p}}{\partial \hat{z}}.
\end{equation}
(Nonlinear inertia and viscous terms only enter at higher orders.)

The oscillatory component of the cross-sectionally integrated continuity equation yields
\begin{equation}\label{averaged_oscillatory_2}
 \frac{\partial \hat{A}}{\partial t}+A_0 \frac{\partial \hat{w}}{\partial z}=0.
\end{equation}
(Here we have neglected steady area contributions of $O(1/\alpha^2 \ell St )\ll 1$ in (\ref{area_expression}) and used the property that $\hat{w}$ is uniform within each cross section.)

Using (\ref{axial_velocity_decomposition}) and (\ref{pressure_decomposition}), the oscillatory components of the upstream pressure condition and downstream flux condition are given respectively by
\begin{equation}\label{oscupstreamb.c}
\hat{p}=0 \qquad \text{at} \qquad z = 0 \qquad \text{and} \qquad \hat{w}=0 \qquad \text{at} \qquad z=1.
\end{equation}
(Due to the presence of Stokes layers located adjacent to the tube wall, which are passive at leading order,  we do not impose a condition on $\hat{w}$ at the tube wall.)

In the flexible section $z \in (z_2,z_2)$, the oscillatory cross-sectional area,  $\hat{A}$, is determined by the solid mechanics of the wall. In the rigid sections $z\in (0,z_1)$ and $z\in(z_2,1)$, there is no change in the cross-sectional area,  so $\hat{A} = 0$.  It is therefore convenient to solve for the oscillatory flow in these three regions separately,  and then match the solutions at the joins $z=z_1,z_2$.  For the fluid mechanics, we require continuity of axial flux and pressure.  Hence, 
\begin{equation}\label{continuity_conditions_osc1}
[\hat{w}]^+_-=[\hat{p}]^+_-=0,\qquad \text{at} \qquad z=z_1,z_2.
\end{equation}
It is also convenient to eliminate $\hat{w}$ from the oscillatory fluid-mechanical equations.  Combining (\ref{oscialltory_inertial_balance}) and (\ref{averaged_oscillatory_2}), we obtain
\begin{equation}\label{second_order_coupling}
\frac{\partial^2 \hat{p}}{\partial z^2}= \frac{1}{A_0}\frac{\partial^2 \hat{A}}{\partial t^2}.
\end{equation}
Equation (\ref{second_order_coupling}) provides a relationship between the oscillatory component of the fluid pressure and oscillatory change to the tube's cross-sectional area at leading order.  In \S\ref{unsteady_coupled_problem_section},  we will use this result to couple the fluid and solid mechanics,  formulating the problem in terms of only the pressure.  

Using (\ref{oscialltory_inertial_balance}),  the full set of oscillatory boundary and matching conditions (\ref{oscupstreamb.c}) and (\ref{continuity_conditions_osc1}) can be written in terms of just $\hat{p}$:
\begin{equation}\label{continuity_conditions}
\hat{p}=0 \qquad \text{at} \qquad z =0,  \qquad \text{and} \qquad \pdiff{\hat{p}}{z} =0 \qquad \text{at} \qquad z=1
\end{equation}
and
\begin{equation}\label{matching_conditions}
\left[\frac{\partial \hat{p}}{\partial z}\right]^+_-=[\hat{p}]^+_-=0 \qquad \text{at} \qquad z=z_1,z_2.
\end{equation}
\subsection{Wall mechanics}\label{wall_mechanics_section}
The external pressure $p^*_{\mathrm{ext}}$ is assumed to be uniform and the fluid mechanics in \S\ref{fluid_mechanics_section} tells us that the internal pressure $p^*$ is uniform in each cross section at leading order.   Hence,  at leading order, the tube is forced by an azimuthally uniform transmural pressure $p_{\mathrm{tm}}^*$.  In response,  it undergoes small-amplitude deformations about its initial elliptical configuration. 

Valid within the parameter regime considered here, \cite{Netherwood2023deformations} used an eigenfunction expansion method to derive a system of equations that collectively describe the small-amplitude displacements of an initially elliptical elastic-walled tube that deforms in response to an applied transmural pressure.  It is this model that we shall adopt here.  They showed that the area perturbation $A^*(z^*,t^*)-A^*_0$ in the flexible section $z_1 L<z^*<z_2L$ could be decomposed into a series of contributions, $A_n^*$,  indexed by the azimuthal mode number, $n$
\begin{equation}\label{A^star_series_expression}
A^*(z^*,t^*)-A_0^*=\sum_{n=1}^\infty A_n^*(z^*,t^*). 
\end{equation}
They showed that each component $A_n^*$ is governed by the partial differential equation
\begin{equation}\label{tubelaw_coupled}
ma\frac{\partial^2}{\partial {t^*}^2}\left(\frac{A_n^*}{A_0^*}\right)-\frac{F}{2 \pi } \frac{\partial^2}{\partial {z^*}^2}\left(\frac{A_n^*}{A_0^*}\right)+\frac{\lambda_n K}{a^3}\left(\frac{A_n^*}{A_0^*}\right)=q_nt_np_{\mathrm{tm}}^*,
\end{equation}
subject to the pinned-end boundary conditions
\begin{equation}\label{dimensional_boundary_conditions}
A_n^* = 0 \qquad \text{at} \qquad z^*=z_1L,z_2L.
\end{equation}
The positive constants $\lambda_n$ and $q_nt_n$ depend only on the initial ellipticity of the tube. They are determined numerically through the solution of an eigenvalue problem.  A detailed explanation leading to the determination of $\lambda_n$ and $q_nt_n$ can be found in \cite{Netherwood2023deformations},  together with tabulated values corresponding to the representative $\sigma_0$ values, as well as continuous plots. 

We decompose each component $A_n^*$ into its respective steady and oscillatory parts using the same scales present in (\ref{area_expression}) for $A^*$:
\begin{align}
A_n^*(z^*,t^*)=a^2 A_n(z,t) = a^2\left(\frac{1}{{\alpha}^2 \ell St}\bar{A}_n(z)+\Delta(t) \hat{A}_n(z,t)\right). \label{area_expression_A_n}
\end{align}
Hence,  by (\ref{area_expression}),(\ref{A^star_series_expression}) and (\ref{area_expression_A_n}),  it follows that
\begin{equation}\label{A_n_A_hat_sum_relationships}
\bar{A}(z)=\sum_{n=1}^\infty \bar{A}_n(z) \qquad \text{and} \qquad \hat{A}(z,t)= \sum_{n=1}^\infty \hat{A}_n(z,t).
\end{equation}

Substituting expressions (\ref{dimensional_transmural}) and  (\ref{area_expression_A_n}) for $p_{\mathrm{tm}}^*$ and $A_n^*$ into (\ref{tubelaw_coupled})--(\ref{dimensional_boundary_conditions}) and making use of the other scalings in \S\ref{set_up_section_coupled}, we find that the non-dimensional governing equations for the steady and oscillatory wall deformations in the flexible part of the tube $z_1<z<z_2$ are given respectively by:
\begin{equation}\label{steady_component_of_tube_law}
\tilde{F}\frac{\D[2]{\bar{A}_n}}{\D{z^2}}-\lambda_n \bar{A}_n=-A_0q_nt_n\left(\bar{p}-\bar{p}_{\mathrm{ext}}\right),
\end{equation}
subject to
\begin{equation}\label{steady_comp_tube_law_bcs}
\bar{A}_n=0 \qquad \text{at} \qquad z=z_1,z_2,
\end{equation}
and
\begin{equation}\label{Oscillatory_component_of_tube_law}
\tilde{F}\frac{\partial^2 \hat{A}_n}{\partial z^2}- M\frac{\partial^2 \hat{A}_n}{\partial t^2} -\lambda_n\hat{A}_n=-A_0q_nt_n \hat{p}(z,t),
\end{equation}
subject to
\begin{equation}\label{dimensionless_oscilatory_b.c}
\hat{A}_n =0 \qquad \text{at} \qquad z=z_1,z_2.
\end{equation}

In the rigid sections of the tube occupying $0<z<z_1$ and $z_2<z<1$, we must ensure that the tube's cross section remains fixed. We therefore impose 
\begin{equation}\label{zero_displacement_rigid}
\bar{A}_n=\hat{A}_n=0 \qquad \text{for} \qquad  z \in (0,z_1)\qquad \text{and} \qquad z \in (z_2,1).
\end{equation}
\section{Oscillatory fluid--structure interaction}\label{unsteady_coupled_problem_section}
From (\ref{area_expression_A_n})--(\ref{A_n_A_hat_sum_relationships}), the steady perturbation to the tube's cross-sectional area is $O(1/\alpha^2 \ell St)$,  which is small for the parameter regime considered here.  As a result of this,  the steady fluid and solid mechanics decouple at leading order,  and the oscillatory fluid mechanics are unaffected by any steady changes to the tube's cross-sectional area.  In the present work,  we focus on using our fluid--structure interaction model to predict the onset of (unsteady) self-excited oscillations in the tube wall.  We therefore proceed by considering only the leading-order oscillatory component of the problem described in \S\ref{set_up_section_coupled}--\S\ref{modelling_section}.  A detailed analysis of the steady problem can be found in the Ph.D. thesis of \citet{Netherwood2024thesis}.  

Our analysis of the oscillatory fluid--structure interaction starts by decomposing the pressure perturbation $\hat{p}$ into modes $\hat{p}_n$ corresponding to the azimuthal area perturbation modes $\hat{A}_n$.  We then eliminate $\hat{A}_n$ in favour of $\hat{p}_n$.  A solution is then obtained in the rigid sections of the tube,  and a governing problem for the fluid--structure interaction within the flexible section is derived.  A solution within the flexible section is posed as a series expansion in $n$,  and the first few terms are obtained. 
\subsection{Formulation in terms of pressure modes}
Motivated by equations (\ref{second_order_coupling})--(\ref{matching_conditions}) and (\ref{A_n_A_hat_sum_relationships}),  we introduce the modal contributions $\hat{p}_n(z,t)$, which are defined as the solution of the following system:
\begin{equation}\label{coupling_p_n_hat}
\frac{\partial^2 \hat{p}_n}{\partial z^2}=\frac{1}{A_0}\frac{\partial \hat{A}_n}{\partial t^2},
\end{equation}
subject to: 
\begin{equation}\label{modal_endpoint_conditions}
\hat{p}_n = 0 \qquad \text{at} \qquad  z=0 \qquad \text{and} \qquad  \pdiff{\hat{p}_n}{z} =0 \qquad \text{at} \qquad z=1, 
\end{equation}
and 
\begin{equation}\label{modal_matching_conditions}
[\hat{p}_n]_{-}^{+} = \left[\pdiff{\hat{p}_n}{z}\right]_{-}^{+}=0 \qquad\text{at} \qquad z = z_1,z_2.  
\end{equation}
With this definition,  $\sum_{n=1}^\infty \hat{p}_n$ satisfies equation (\ref{second_order_coupling}) for $\hat{p}$,  and the boundary and matching conditions (\ref{continuity_conditions})--(\ref{matching_conditions}) for $\hat{p}$. Therefore, we deduce that 
\begin{equation}\label{pressure_into_p_n}
\hat{p}(z,t)=\sum_{n=1}^\infty \hat{p}_n(z,t).
\end{equation}
Hence, we interpret $\hat{p}_n$ as the component of the transmural pressure corresponding to the $n$th azimuthal mode of the wall deformations.
\subsection{Solution in the rigid sections}
In the rigid sections, the tube's cross section is fixed. Using (\ref{zero_displacement_rigid}) and (\ref{coupling_p_n_hat}), we derive the governing equation 
\begin{equation}\label{rigid_governing_equation}
\frac{\partial^2 \hat{p}_n}{\partial z^2}=0, 
\end{equation}
valid for $0<z<z_1$ and $z_2<z<1$.

Solving (\ref{rigid_governing_equation}) subject to (\ref{modal_endpoint_conditions}), we find that 
\begin{equation}\label{solution_rigid_p_hat}
  \hat{p}_n(z,t)=\begin{cases}
    G_n(t)z, & \qquad \text{for} \qquad z \in (0,z_1),\\
    H_n(t), & \qquad \text{for} \qquad z \in (z_2,1),
  \end{cases}
\end{equation}
where $G_n(t)$ and $H_n(t)$ are arbitrary functions of time,  which will later be determined during the process of matching solutions between rigid and flexible sections. 
\subsection{Governing equations and boundary conditions for $\hat{p}_n$ in the flexible section}\label{flexible_equations_section}
Eliminating $\hat{A}_n$ between (\ref{Oscillatory_component_of_tube_law}) and (\ref{coupling_p_n_hat}), the governing equation for $\hat{p}_n$,  which governs the oscillatory fluid--structure interaction in the flexible section of the tube is:
\begin{equation}\label{governing_eq_p_n_hat}
\tilde{F}\frac{\partial^4 \hat{p}_n}{\partial z^4}-M\frac{\partial^4 \hat{p}_n}{\partial t^2 \partial z^2}-\lambda_n\frac{\partial^2 \hat{p}_n}{\partial z^2}=-q_nt_n \sum_{i=1}^\infty \frac{\partial^2 \hat{p}_i}{\partial t^2}.
\end{equation}
Substituting the solution (\ref{solution_rigid_p_hat}) into the four matching conditions (\ref{modal_matching_conditions}),  and eliminating $G_n$ and $H_n$,  we obtain the following explicit boundary conditions on $\hat{p}_n$:
\begin{align}
z_1\frac{\partial \hat{p}_n}{\partial z}-\hat{p}_n&=0, \qquad \text{at} \qquad z=z_1, \label{z_1_join_condition} \\
\frac{\partial \hat{p}_n}{\partial z}& =0, \qquad \text{at} \qquad z=z_2. \label{z_2_join_condition}
\end{align}
Combining (\ref{dimensionless_oscilatory_b.c}) with (\ref{coupling_p_n_hat}),  we also need
\begin{equation}\label{fixed_area_p_n_hat_cond}
\frac{ \partial^2 \hat{p}_n}{\partial z^2}=0 \qquad \text{at} \qquad z=z_1,z_2.
\end{equation}
The governing equations (\ref{governing_eq_p_n_hat}) and their boundary conditions (\ref{z_1_join_condition})--(\ref{fixed_area_p_n_hat_cond}) are linear and homogeneous in the $\hat{p}_i$.  The pressure will therefore have arbitrary amplitude,  which we shall set by imposing the normalisation 
\begin{equation}\label{normalisation1_p_n_hat}
 \int_{0}^{1}\bigg(\pdiff{\hat{p}}{z}\bigg)^2\D{z}  = z_1\left(\sum_{n=1}^\infty \pdiff{\hat{p}_n}{z}\bigg|_{z=z_1}\right)^2+ \int_{z_1}^{z_2}\bigg(\frac{\partial }{\partial z}\sum_{n=1}^\infty \hat{p}_n\bigg)^2\text{d} z = 1.
\end{equation}
The boundary-value problem for the $\hat{p}_n$ consists of a coupled set of linear homogeneous fourth-order partial differential equations (\ref{governing_eq_p_n_hat}),  together with four sets of boundary conditions (\ref{z_1_join_condition})--(\ref{z_2_join_condition}) and (\ref{fixed_area_p_n_hat_cond}) as well as the global normalisation condition (\ref{normalisation1_p_n_hat}).  

It is convenient to introduce the scaled axial co-ordinate
\begin{equation}\label{scaled_axial}
\zeta= \frac{z_2-z}{z_2-z_1},
\end{equation}
and seek solutions in which $\hat{p}_n$ varies harmonically in time with dimensionless frequency $\omega$.  We therefore introduce $\tilde{p}_n$ and $\tilde{p}$ and write
\begin{equation}\label{p_n_harmonic}
\hat{p}_n(z,t) = Re\left(\tilde{p}_n(\zeta) e^{i\omega t} \right) \qquad \text{and} \qquad \hat{p}(z,t) = Re \left(\tilde{p}(\zeta) e^{i\omega t}\right).
\end{equation}
With these definitions,  it follows from (\ref{pressure_into_p_n}) that $\tilde{p} = \sum_{n=1}^\infty \tilde{p}_n$.

Substituting (\ref{p_n_harmonic}) into (\ref{z_1_join_condition})--(\ref{normalisation1_p_n_hat}), the problem for $\tilde{p}_n$ is given by
\begin{align}\label{flexible_governing}
\frac{\D[4]{\tilde{p}_n}}{\D{\zeta^4}}+\frac{1}{\tilde{F}}\left(M\omega^2 -\lambda_n\right)\left(z_2-z_1\right)^2\frac{\D[2]{\tilde{p}_n}}{\D{\zeta^2}}=\frac{\omega^2 q_n t_n}{\tilde{F}}\left(z_2-z_1\right)^4\sum_{i=1}^\infty \tilde{p}_i(\zeta), 
\end{align}
subject to the boundary conditions
\begin{align}
\frac{\D{\tilde{p}_n}}{\D{\zeta}}&=0 \;\;\; \text{at} \;\;\; \zeta=0, \\
\frac{z_1}{z_2-z_1}\frac{\D{\tilde{p}_n}}{\D{\zeta}}+\tilde{p}_n &=0 \;\;\; \text{at}\;\;\; \zeta=1, \\
\frac{\D[2]{\tilde{p}_n}}{\D{\zeta^2}}&=0 \;\;\; \text{at}\;\;\; \zeta=0,1, \label{total_bc_p_n_3}
\end{align}
and the normalisation
\begin{equation}\label{normlisation_zeta}
\frac{z_1}{(z_2-z_1)^2}\left(\sum_{n=1}^\infty \frac{\D{\tilde{p}_n}}{\D{\zeta}}\bigg|_{\zeta = 1}\right)^2+\frac{1}{z_2-z_1}\int_0^1 \left(\sum_{n=1}^\infty \frac{\D{\tilde{p}_n}}{\D{\zeta}} \right)^2 \D{\zeta}=1.
\end{equation}
\subsection{Series solution for $\tilde{p}$ in the flexible section}\label{flexible_section_series_section}
\cite{Netherwood2023deformations} found that the numerical constants $q_nt_n$ decay rapidly as $n$ is increased.  From (\ref{flexible_governing}),  we see that this means the forcing for $\tilde{p}_1$ will be larger than $\tilde{p}_2$,  and larger still than $\tilde{p}_3$,  etc.  Hence,  we expect that $\tilde{p}_1$ will provide the dominant contribution to the global solution for the pressure,  with each successive mode $\tilde{p}_i$ being smaller as the mode number $i$ increases.  

We investigate the large-$n$ behaviour of $q_nt_n$ in Appendix \ref{asymptotic_appendix} by fitting curves to the numerical data obtained by \cite{Netherwood2023deformations}.  Our analysis suggests the asymptotic behaviour 
\begin{equation}
q_nt_n \sim Q \epsilon^{n-1} \qquad \text{as} \qquad n \to \infty,
\end{equation}
where $Q$ and $\epsilon \in (0,1)$ are constants that depend on $\sigma_0$. 

Having estimated $\epsilon$ (see Appendix \ref{asymptotic_appendix}),  we introduce $Q_n$ such that 
\begin{equation}\label{asymptotic_axiom}
q_nt_n = Q_n \epsilon^{n-1}. 
\end{equation}
Hence,  $Q_n \to Q$ as $n \to \infty$.  

The $Q_n$ are found to decrease monotonically with increasing $n$.  We therefore consider a formal power series expansion in $\epsilon$,  assuming $q_nt_n = O(\epsilon^{n-1})$.  

We expand $\omega$ and $\tilde{p}_n$ in powers of $\epsilon$:
\begin{align}
\omega &= \omega_0+\epsilon\omega_1 +\epsilon^2 \omega_2+\epsilon^3 \omega_3 + O(\epsilon^4), \label{asymptotic_omega} \\
\tilde{p}_n &=\epsilon^{n-1}\left(p_{n0}+\epsilon p_{n1}+\epsilon^2 p_{n2} + \epsilon^3 p_{n3}+O(\epsilon^4)\right),\label{asymptotic_pressure}
\end{align}
where the pre-factor $\epsilon^{n-1}$ present in (\ref{asymptotic_pressure}) has been chosen after observing how the scaling in (\ref{asymptotic_axiom}) affects the right-hand-side of (\ref{flexible_governing}). 

From (\ref{p_n_harmonic}) and (\ref{asymptotic_pressure}),  the full axial mode for the pressure is given by 
\begin{align}
\tilde{p}(z) &= \sum_{n=1}^\infty \tilde{p}_n(\zeta) =  p_{10}+\epsilon(p_{11}+p_{20}) + \epsilon^2(p_{12} + p_{21} + p_{30})+O(\epsilon^3).\label{pressure_expansion}
\end{align}

We now substitute (\ref{asymptotic_omega})--(\ref{pressure_expansion}) into (\ref{flexible_governing})--(\ref{normlisation_zeta}) and equate at increasing powers of $\epsilon$.  The problem at $O(\epsilon^0)$ involves only the first azimuthal mode ($n=1$),   and defines an eigenvalue problem for $p_{10}$ and $\omega_0$.  At $O(\epsilon)$,  only the $n=1$ and $n=2$ modes will contribute,  which yields a problem for $p_{20}$,  and a separate eigenvalue problem for $p_{11}$ and $\omega_1$.  The system for $p_{20}$ is forced by $p_{10}$ and $\omega_0$,  and the system for $p_{11}$ is forced by $p_{20}$.  At $O(\epsilon^2)$,  only the $n=1,2,3$ modes will contribute,  yielding problems for $p_{21}$ and $p_{30}$, which then force a new eigenvalue problem for $p_{12}$ and $\omega_2$.  

This pattern continues as we equate at higher orders of magnitude in $\epsilon$.  A series solution can therefore be found by computing the relevant components iteratively, starting with the $O(\epsilon^0)$ problem for $p_{10}$ and $\omega_0$.  
\subsection{The $O(\epsilon^0)$ problem for $\omega_0$ and $p_{10}$}\label{leading_order_section}
Substituting (\ref{asymptotic_omega})--(\ref{asymptotic_pressure}) into (\ref{flexible_governing})--(\ref{normlisation_zeta}) and equating at $O(\epsilon^0)$ with $n=1$, we find that $p_{10}(\zeta)$ and $\omega_0$ satisfy the following eigenvalue problem:
\begin{equation}\label{p_10_scaled_governing}
\frac{\D[4]{p_{10}}}{\D{\zeta^4}}+\frac{1}{\tilde{F}}\bigg((M\omega_0^2-\lambda_1)(z_2-z_1)^2\bigg)\frac{\D[2]{p_{10}}}{\D{\zeta^2}}-\frac{1}{\tilde{F}}Q_1\omega_0^2(z_2-z_1)^4p_{10}=0,
\end{equation}
subject to:
\begin{align}
\frac{\D{p_{10}}}{\D{\zeta}}&=0 \;\;\; \text{at} \;\;\; \zeta =0, \label{p_10_scaled_bc_1} \\
\frac{z_1}{z_2-z_1}\frac{\D{p_{10}}}{\D{\zeta}}+p_{10}&=0 \;\;\; \text{at} \;\;\; \zeta =1, \label{p_10_scaled_bc_2} \\
\frac{\D[2]{p_{10}}}{\D{\zeta^2}}&=0 \;\;\; \text{at} \;\;\; \zeta =0,1, \label{p_10_scaled_bc_3}
\end{align}
and the normalisation
\begin{equation}\label{leading_order_normalisation}
\frac{z_1}{(z_2-z_1)^2}\left(\frac{\D{p_{10}}}{\D{\zeta}}\right)^2\bigg|_{\zeta =1} + \frac{1}{z_2-z_1}\int_0^1 \bigg(\frac{\D{p_{10}}}{\D{\zeta}}\bigg)^2\D{\zeta} = 1.
\end{equation}

The eigenvalue problem (\ref{p_10_scaled_governing})--(\ref{leading_order_normalisation}) is of the same form as the problem derived by \citet{walters2018effect} (and \citet{whittaker2010predicting} for the case $M=0$).  However,  it differs through the values of the numerical constants $\lambda_1$ and $Q_1=q_1t_1$. We proceed using the same solution method, although we omit some of the detailed calculations since they are covered comprehensively by \citet{walters2018effect} and \citet{whittaker2010predicting}.

The general solution to (\ref{p_10_scaled_governing}) for $p_{10}$ can be written as
\begin{equation}\label{general_solution_p_10}
p_{10}(\zeta)=A\frac{\sinh(g\zeta)}{\sinh(g)}+ B\frac{\sinh(g(1- \zeta))}{\sinh(g)}+C\cos(h \zeta)+D \sin(h \zeta),
\end{equation}
where $A,B,C$ and $D$ are arbitrary constants, and $g$ and $h$ are real positive constants given by
\begin{equation}\label{def_g}
g^2 = (z_2-z_1)^2\left[-\frac{(M \omega_0^2-\lambda_1)}{2 \tilde{F}}+\sqrt{\frac{(M\omega_0^2 -\lambda_1)^2}{4\tilde{F}^2}+\frac{1}{\tilde{F}}Q_1\omega_0^2 }\right],
\end{equation}
and
\begin{equation}\label{def_h}
h^2 = (z_2-z_1)^2\left[\frac{M \omega_0^2-\lambda_1}{2 \tilde{F}}+\sqrt{\frac{(M\omega_0^2 -\lambda_1)^2}{4\tilde{F}^2}+\frac{1}{\tilde{F}}Q_1\omega_0^2 }\right].
\end{equation}
\subsubsection{Solution for $\omega_0$}\label{sol_for_omega_0_section}
Substituting the general solution (\ref{general_solution_p_10}) into the boundary conditions (\ref{p_10_scaled_bc_1})--(\ref{p_10_scaled_bc_3}) and seeking only non-trivial solutions, the following eigenvalue equation for $\omega_0$ is obtained
\begin{align}\label{leading_order_eigenvalue_equation}
z_1\Bigl[2gh(&1-\cosh(g)\cos(h))+(g^2-h^2)\sinh(g)\sin(h)\Bigr]\nonumber \\&-(z_2-z_1)\frac{g^2+h^2}{gh}\Bigl[g\sinh(g)\cos(h)+h\cosh(g)\sin(h)\Bigr]=0.
\end{align}

In total there are three equations (\ref{def_g})--(\ref{leading_order_eigenvalue_equation}) relating $g,h$ and $\omega_0$. Following \citet{walters2018effect}, we take the difference and product of (\ref{def_g})--(\ref{def_h}) in turn, yielding
\begin{equation}\label{omega_0_formula}
\omega_0^2=\frac{\tilde{F}g^2 h^2}{Q_1(z_2-z_1)^4},
\end{equation}
and
\begin{equation}\label{g_in_terms_of_h}
g=\left[\frac{\frac{\lambda_1}{\tilde{F}}(z_2-z_1)^2+h^2}{1+\frac{M h^2}{Q_1 (z_2-z_1)^2}}\right]^{1/2}.
\end{equation}

Observe that (\ref{g_in_terms_of_h}) allows for the elimination of $g$ from (\ref{leading_order_eigenvalue_equation}). The resulting equation for $h$ can then be solved numerically. Having obtained solutions for $h$, corresponding values of $g$ and $\omega_0$ can be recovered using (\ref{g_in_terms_of_h}) and (\ref{omega_0_formula}) in turn. 

Consistent with the results of \citet{walters2018effect}, we find countably many solutions for $h$ (and hence $g$ and $\omega$), which we shall index with an oscillatory mode number,  $j$.  We denote $\omega_{0}^{(1)}$ as the fundamental (lowest) oscillation frequency. In table \ref{eigenfrequencies} we tabulate the first five modes of the leading-order oscillation frequency $\omega_0^{(j)}$ for different values of $M$ and $\tilde{F}$, with $\sigma_0 = 0.6$. 
\begin{figure}
\centering
\includegraphics[scale=0.37]{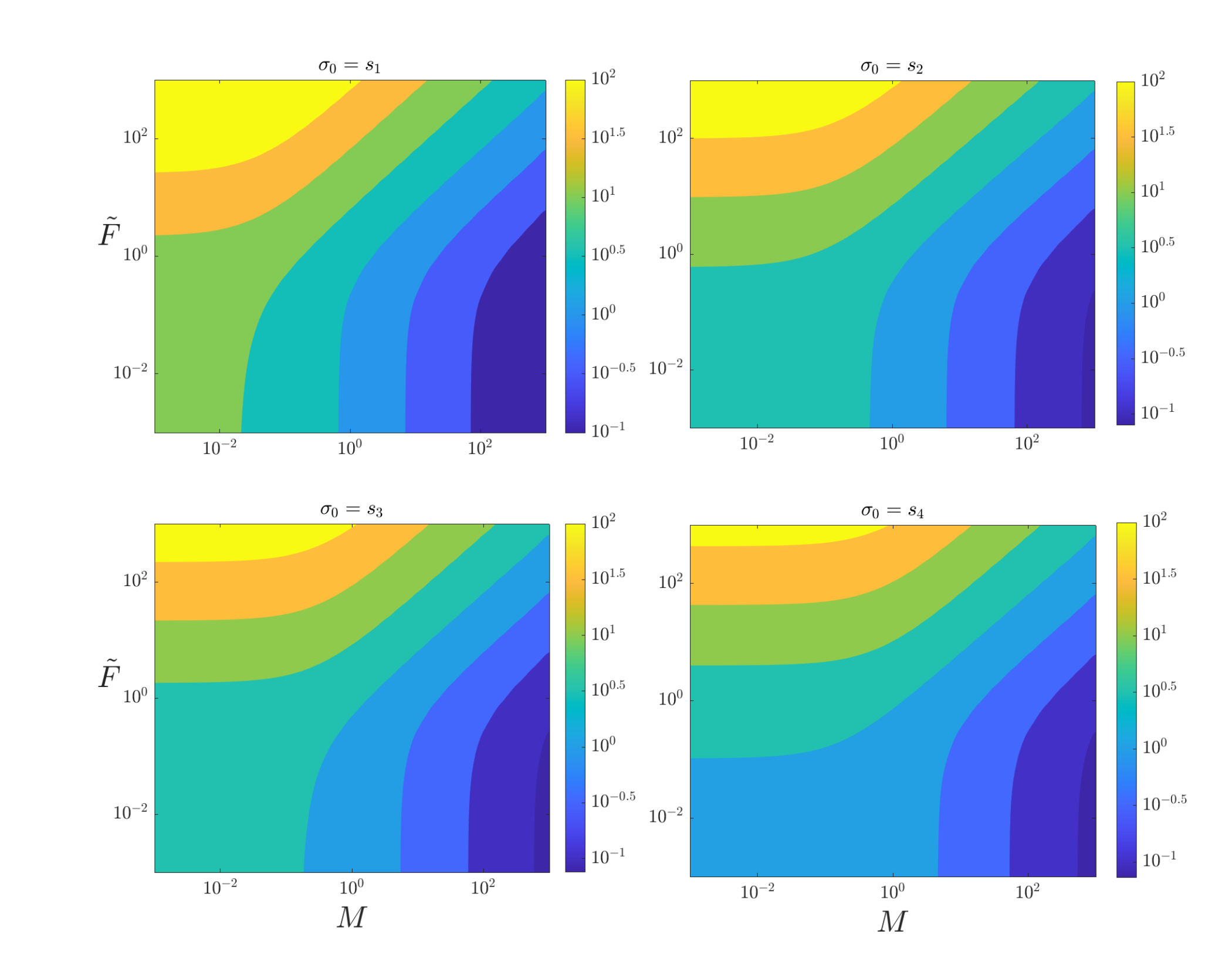}
\caption[Numerical solutions for the leading-order oscillation frequencies $\omega_0^{(1)}$.]{Values of $\omega_0^{(1)}$ plotted throughout $(M, \tilde{F})$ parameter space, for $\sigma_0 = s_1, s_2, s_3, s_4$. The values were calculated by substituting the numerical solutions for $g$ and $h$ into (\ref{omega_0_formula}), with $z_1 = 0.1, z_2 = 0.9$. The values for $Q_1=q_1t_1$ were obtained from the numerical data for $q_nt_n$,  which can be found in \cite{Netherwood2023deformations}.}\label{omega_0_contour_plot}
\end{figure}

In figure \ref{omega_0_contour_plot},  we provide contour plots of the values for $\omega_0^{(1)}$ as $M$ and $\tilde{F}$ are varied at different values of $\sigma_0$.  In general,  we find that the oscillation frequency increases with $\tilde{F}$ and decreases with $M$. As we vary $\sigma_0$, we find that tubes with an initially almost circular cross section (larger $\sigma_0)$ have larger natural eigenfrequencies than those with initially larger ellipticities (smaller $\sigma_0$), particularly for large $\tilde{F}$.  
\begin{table}
\centering
\begin{tabular}{r| r| r r r r r r}
\hline
$\;\; \tilde{F}$ & $\;\;M$     & $\;\;\omega_{0}^{(1)}$ & $\;\;\omega_{0}^{(2)}$ & $\;\;\omega_{0}^{(3)}$ & $\;\;\omega_{0}^{(4)}$ & $\;\;\omega_{0}^{(5)}$   \\    \hline 
{}& 0       & $6.119$       & $19.104$      & $33.881$      & $51.012$      & $70.921$      \\
{}& $0.001$ & $6.102$       & $18.637$      & $31.660$      & $44.876$      & $58.056$      \\ 
0.01& $0.01$  & $5.956$       & $15.549$      & $21.584$      & $25.679$      & $29.076$      \\ 
{}& $0.1$   & $4.908$       & $7.697$       & $8.528$       & $9.239$       & $9.999$       \\ 
{}& $1$     & $2.379$       & $2.626$       & $2.779$       & $2.974$       & $3.199$      \\ \hline
{}& 0       & $7.010$       & $25.73$       & $54.591$      & $94.925$      & $147.457$     \\ 
{}& $0.001$ & $6.989$       & $25.176$      & $50.901$      & $83.313$      & $120.488$     \\ 
0.1& $0.01$  & $6.811$       & $20.854$      & $34.449$      & $47.566$      & $60.397$      \\ 
{}& $0.1$   & $5.545$       & $10.224$      & $13.620$      & $17.220$      & $20.852$      \\ 
{}& $1$     & $2.629$       & $3.504$       & $4.474$       & $5.559$       & $6.689$       \\ \hline
{}& 0       & $11.865$      & $59.792$      & $145.235$     & $269.936$     & $434.357$     \\ 
{}& $0.001$ & $11.828$      & $58.238$      & $135.356$     & $236.860$     & $354.879$     \\ 
1& $0.01$  & $11.509$      & $48.152$      & $91.528$      & $135.323$     & $178.012$     \\ 
{}& $0.1$   & $9.284$       & $23.641$      & $36.295$      & $49.092$      & $61.520$      \\ 
{}& $1$     & $4.357$       & $8.145$       & $11.961$      & $15.859$      & $19.745$      \\  
\end{tabular}
\caption[Numerical solutions for the leading-order oscillation frequencies $\omega_0^{(1)}$ .]{Values for the oscillation frequencies $\omega_{0}^{(j)}$, tabulated for $j=1,2,\cdots,5$ using $\sigma_0 = 0.6,z_1=0.1, z_2=0.9$, and a range of $(M, \tilde{F})$. To obtain the values we substituted the numerical solutions for $g$ and $h$, as well the numerical data for $Q_1$ into the formula (\ref{omega_0_formula}).  The values for $Q_1=q_1t_1$ were obtained from the numerical data for $q_nt_n$,  which can be found in \cite{Netherwood2023deformations}.}\label{eigenfrequencies}
\end{table}
\subsubsection{Solution for $p_{10}$}\label{sol_p_10_section}
Substituting the general solution (\ref{general_solution_p_10}) into the boundary conditions (\ref{p_10_scaled_bc_1}) and (\ref{p_10_scaled_bc_3}), we can derive analytical expressions for the constants $A,B$ and $C$ in terms of $g$, $h$ and the final constant, $D$. The final boundary condition (\ref{p_10_scaled_bc_2}) is satisfied automatically due to $\omega_0$ being chosen to satisfy the eigenvalue equation (\ref{leading_order_eigenvalue_equation}). We find that 
\begin{align}
A & =\left[ -\frac{h \left(\cos\! \left(h\right) \sinh\! \left(g\right) g+\sin\! \left(h\right) h \cosh\! \left(g\right)\right)}{g^{2} \left(\cos\! \left(h\right)-\cosh\! \left(g\right)\right)}\right]D,\label{A_p10}\\
B & = \left[-\frac{h \left(\sin\! \left(h\right) h+\sinh\! \left(g\right) g\right)}{g^{2} \left(\cos\! \left(h\right)-\cosh\! \left(g\right)\right)}\right]D,\\
C & =\left[\frac{ \sin\! \left(h\right) h+\sinh\! \left(g\right) g}{h \left(\cos\! \left(h\right)-\cosh\! \left(g\right)\right)}\right]D,\label{C_p10}
\end{align}
where the parameters $g$ and $h$ are given in terms of $\omega_0$ and the other problem parameters by (\ref{def_g}) and (\ref{def_h}) respectively. The final constant, $D$, is set using the normalisation condition (\ref{leading_order_normalisation}).  In principle,  an analytical expression can be obtained for $D$ in terms of $g$, $h$ and $\omega_0$.  However,  this expression is prohibitively complex,  so we choose not to present it here.   We define $p_{10}^{(j)}$ as the solution of (\ref{p_10_scaled_governing})--(\ref{leading_order_normalisation}) corresponding to the $j$th oscillatory mode.

\begin{figure}
\centering 
\includegraphics[scale=0.37]{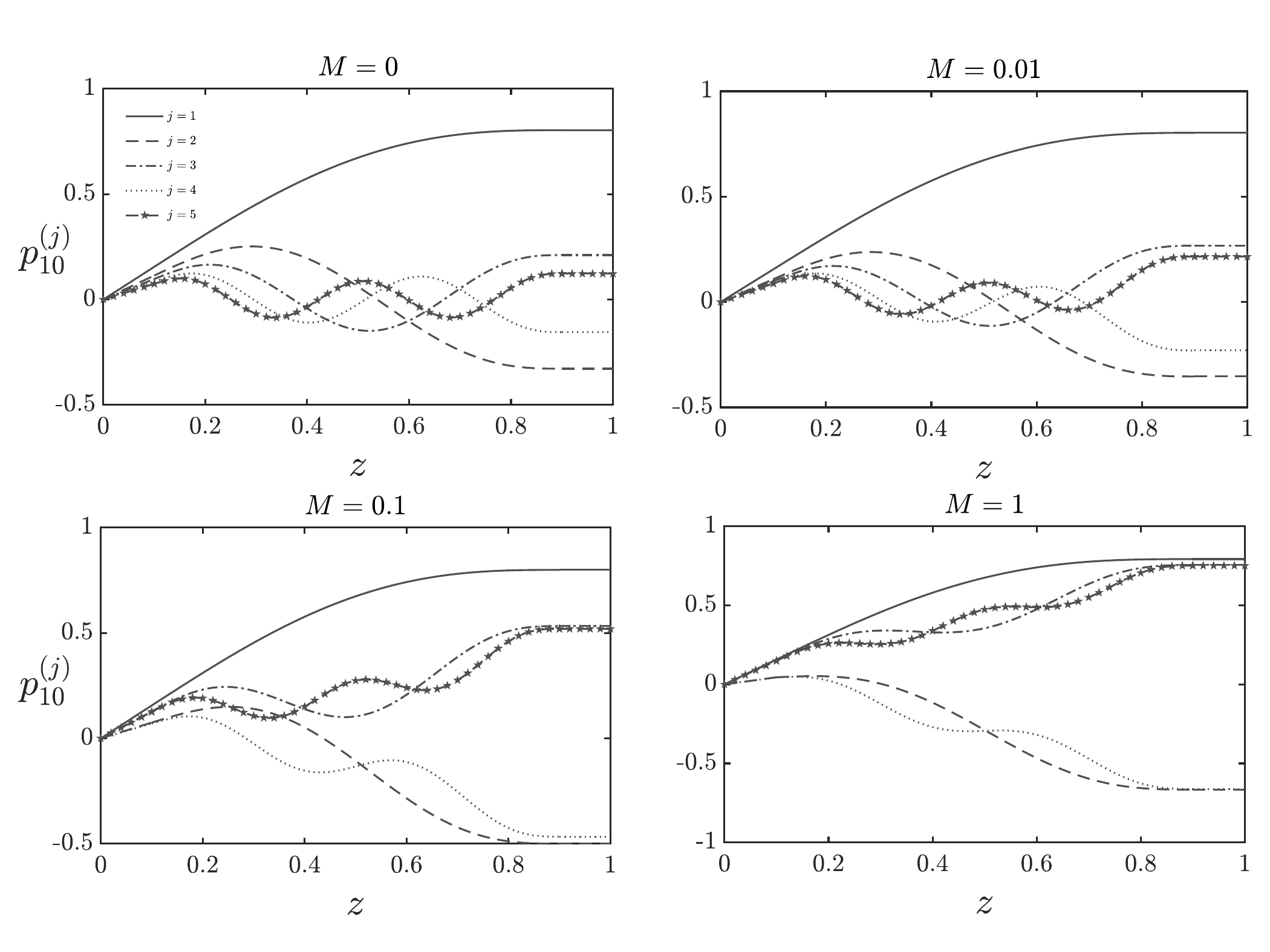}
\caption[Solutions for $p_{10}^{(j)}(z)$.  ]{Solutions for $p_{10}^{(j)}(z)$ with $z_1=0.1,  z_2=0.9,  \sigma _0 = 0.6,  \tilde{F} =1$,  and different values of the inertia parameter $M$ and mode number $j$. The solutions were obtained by substituting the numerical data for $g$ and $h$ into the analytical solution (\ref{general_solution_p_10}) with (\ref{A_p10})--(\ref{C_p10}) for $p_{10}^{(j)}$.}\label{p_10_solution}\label{solution_for_p10}
\end{figure}

We plot our solutions for $p_{10}^{(j)}$ in figure \ref{solution_for_p10}. In agreement with the results of \citet{whittaker2010predicting} and \citet{walters2018effect},  for the parameters considered here,  we find that the number of extrema present in the solution for $p_{10}^{(j)}$ is equal to the mode number, $j$.  For $j=1$,  increasing the inertia parameter $M$ from $0$ to $1$ has only a small effect on $p_{10}^{(j)}$.  However,  for $j \geq 2$ there are noticeable changes in the curves with $p_{10}^{(j)}$ becoming more positive for odd $j$ and more negative for even $j$. 

For every oscillatory mode $j$,  there is an associated solution of the full coupled problem. \citet{whittaker2010predicting} deduced that unsteady perturbations to the steady problem oscillating with a fundamental $(j=1)$ oscillation frequency were most unstable. Since our interests lie in predicting the onset of self-excited oscillations, for the remainder of this work we present solutions corresponding to only the fundamental $(j=1)$ mode. We proceed by dropping the superscript $(j)$ notation with $j=1$ assumed.
\subsection{The $O(\epsilon)$ problem for $p_{20}$}\label{p_20_section}
Substituting (\ref{asymptotic_omega})--(\ref{asymptotic_pressure}) into (\ref{flexible_governing})--(\ref{normlisation_zeta}) with $n=2$ and equating at $O(\epsilon)$ yields the following system for $p_{20}$:
\begin{equation}\label{p_20_zeta_governing}
\frac{\D[4]{p_{20}}}{\D{\zeta^4}}-\psi^2\frac{\D[2]{p_{20}}}{\D{\zeta^2}}=\frac{1}{\tilde{F}}\omega_0^2Q_2(z_2-z_1)^4p_{10}(\zeta),
\end{equation}
subject to:
\begin{align}
\frac{\D{p_{20}}}{\D{\zeta}}&=0 \qquad \text{at} \qquad \zeta =0,  \label{p_20_bc1_zeta}\\
\frac{z_1}{z_2-z_1}\frac{\D{p_{20}}}{\D{\zeta}}+p_{20}&=0 \qquad \text{at} \qquad \zeta =1, \\
\frac{\D[2]{p_{20}}}{\D{\zeta^2}}&=0 \qquad \text{at} \qquad \zeta =0,1, \label{p_20_bc3_zeta}
\end{align}
where 
\begin{equation}\label{psi_squared_def}
\psi^2=\frac{1}{\tilde{F}}\left(\lambda_2-M\omega_0^2\right)\left(z_2-z_1\right)^2.
\end{equation}

We observe that the governing equation (\ref{p_20_zeta_governing}) for $p_{20}$ is a linear ordinary differential equation with constant coefficients.  The forcing on the right hand side is known,  since $p_{10}(\zeta)$ has already been found.
\subsubsection{Solution for $p_{20}$}
The system (\ref{p_20_zeta_governing})--(\ref{p_20_bc3_zeta}) can be solved for $p_{20}$ using the standard approach of summing a complimentary function and particular integral.  In order to construct the particular integral for $p_{20}$, we must first determine whether the system is resonant, i.e.  whether the forcing from $p_{10}$ coincides with the complementary function for the homogeneous problem.  Given the form (\ref{general_solution_p_10}) of $p_{10}$ and the sign of $\psi^2$, such resonance will occur precisely when $g^2 = \psi^2$.  Hence,  there is a single line in the parameter space $(M,\tilde{F})$ on which (\ref{p_20_zeta_governing})--(\ref{p_20_bc3_zeta}) becomes resonant. For this special case, a general solution can be obtained by assuming that the particular integral for $p_{20}$, takes the form
\begin{equation}
F_1\zeta\frac{\sinh(g\zeta)}{\sinh(g)}+F_2\zeta \frac{\sinh(g(1-\zeta))}{\sinh(g)}+F_3\cos(h\zeta)+F_4\sin(h\zeta),
\end{equation}
where $F_1,F_2,F_3$ and $F_4$ are constants.

For brevity,  we only present in detail the case $\psi^2 \neq g^2$,  though it is possible to obtain solutions when $\psi^2 = g^2$.  For the case $\psi^2 \neq g^2$,  we find that the general solution of (\ref{p_20_zeta_governing})--(\ref{p_20_bc3_zeta}) is given by
\begin{align}\label{p_20_solution}
 p_{20}(\zeta) &= \hat{A}+\hat{B}\zeta +\hat{C} \frac{\sinh(\psi \zeta)}{\sinh(\psi)} + \hat{D} \frac{\sinh(\psi(1- \zeta))}{\sinh(\psi)} \nonumber \\ & \qquad +C_1 \frac{\sinh(g\zeta)}{\sinh(g)}+C_2 \frac{\sinh(g(1-\zeta))}{\sinh(g)}+C_3\cos(h\zeta)+C_4\sin(h\zeta),
\end{align}
where $\hat{A},\hat{B},\hat{C},\hat{D}$ are arbitrary constants in the complementary function, and the constants in the particular integral are given by:
\begin{align}
C_1 & =\frac{\omega_0^2 Q_2}{\tilde{F}}\frac{(z_2-z_1)^4}{g^4(g^2-\psi^2)}\left[\frac{h(g \cos(h) \sinh(g) + h \sin(h)\cosh(g))}{\cosh(g) - \cos(h)} \right]D,\\ 
C_2 &=\frac{\omega_0^2 Q_2}{\tilde{F}}\frac{(z_2-z_1)^4}{g^4(g^2-\psi^2)}\left[\frac{h(h \sin(h) +g \sinh(g))}{\cosh(g)-\cos(h)} \right]D,\\
C_3 & = \frac{\omega_0^2 Q_2}{\tilde{F}} \frac{(z_2-z_1)^4}{h^3(h^2+\psi^2)}\left[ \frac{h \sin(h)+g \sinh(g)}{\cos(h)-\cosh(g)}\right]D, \\
C_4 & = \frac{\omega_0^2 Q_2}{\tilde{F}}\frac{(z_2-z_1)^4}{h^2(h^2+\psi^2)}D, \label{C_4}
\end{align}
where $D$ is the (known) normalisation constant in $p_{10}$.

To determine $\hat{A},\hat{B},\hat{C}$ and $\hat{D}$ we substitute (\ref{p_20_solution})--(\ref{C_4}) into the four boundary conditions (\ref{p_20_bc1_zeta})--(\ref{p_20_bc3_zeta}) and solve the resulting linear system using $\texttt{Maple}$.  Analytical expressions were obtained for $\hat{A},\hat{B},\hat{C}$ and $\hat{D}$ in terms of $z_1,z_2,  M,  \tilde{F}$ and the numerically determined constants $h,g,\omega_0,Q_1$ and $Q_2$.  However,  due to the length of the symbolic expressions,  we choose to omit the expressions here.

\begin{figure}
\centering
\includegraphics[scale=0.25]{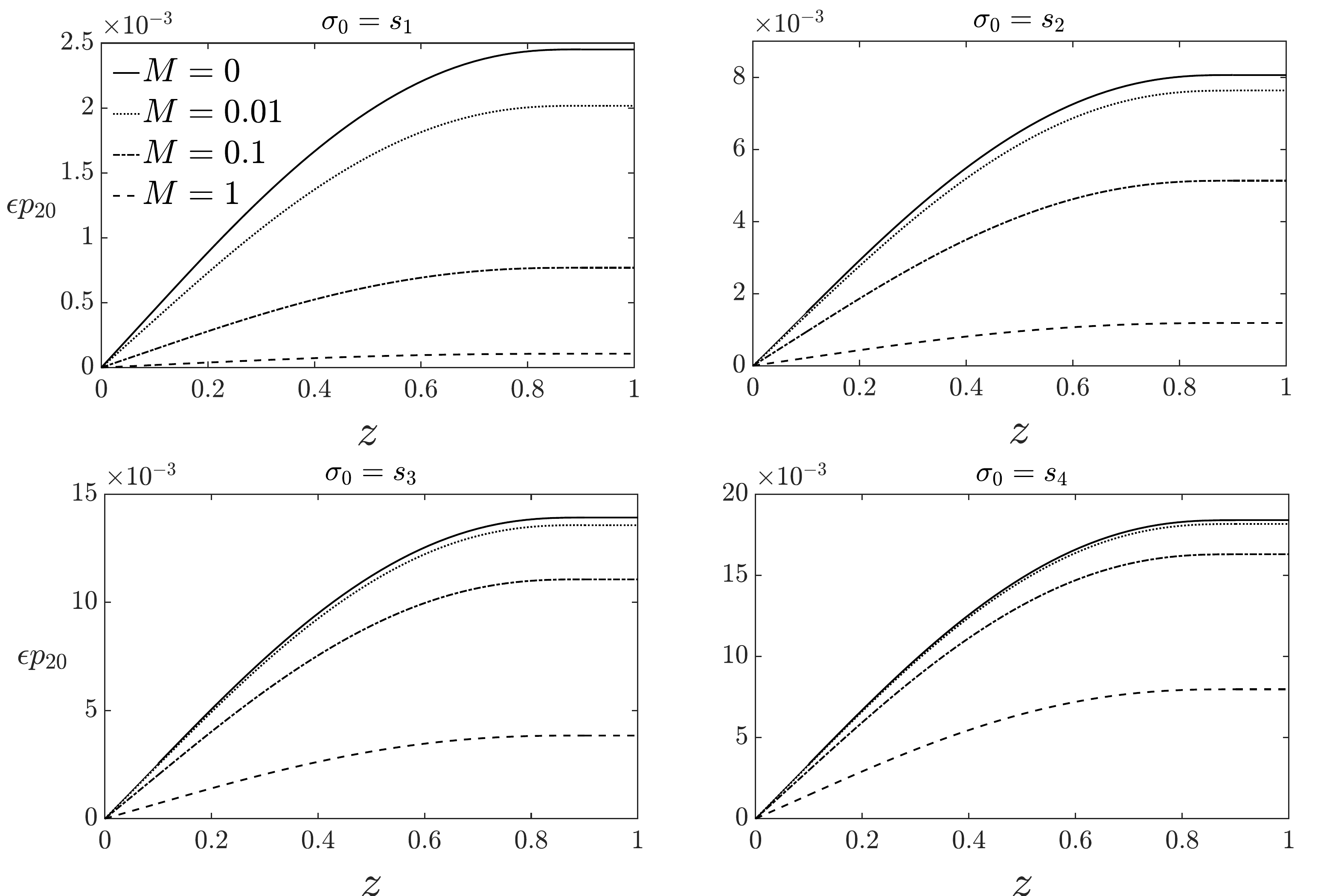}
\caption[Solutions for $\epsilon p_{20}$.]{Solutions for $\epsilon p_{20}$, plotted for $M=0,0.01,0.1,1$ with $\tilde{F}=1$, and $\sigma_0= s_1,s_2,s_3,s_4$ with $z_1=0.1$ and $z_2=0.9$. The solutions were calculated using the expression (\ref{p_20_solution}). The coefficients present in the solutions were determined analytically in terms of the numerical constants $h,g$ and $\omega$ within \texttt{Maple}. The second panel corresponds to $\sigma_0=0.6$ and allows for a comparison with \citet{walters2018effect} and \citet{whittaker2010predicting}. The values for $\epsilon Q_2 = q_2t_2$ were obtained from the numerical data for $q_nt_n$,  which can be found in \cite{Netherwood2023deformations}.}\label{p_20_solution_figure}
\end{figure}

In figure \ref{p_20_solution_figure},  we plot solutions for $\epsilon p_{20}$, which is independent of $\epsilon$, using (\ref{p_20_solution}). Unlike the solutions for $p_{10}$, we find that changes in the inertia coefficient result in an observable difference in the corresponding profiles for $p_{20}$.  We see that $M=0$ results in a solution for $p_{20}$ with maximal amplitude, with the amplitude of the solution then monotonically decreasing as $M$ is increased.  As we vary the ellipticity parameter, our results demonstrate that smaller $\sigma_0$ results in a larger amplitude of the solution.
\subsection{The $O(\epsilon)$ problem for ${p_{11}}$ and $\omega_1$}\label{p_11_section}
Substituting (\ref{asymptotic_omega})--(\ref{asymptotic_pressure}) into (\ref{flexible_governing})--(\ref{normlisation_zeta}) with $n=1$ and equating at $O(\epsilon)$,  we obtain the following problem for $p_{11}$:
\begin{align}\label{p_11_governing}
\mathcal{L}(p_{11})= S, 
\end{align}
where
\begin{equation}
\mathscr{L}(y)=\frac{\D[4]{y}}{\D{\zeta^4}}+P\frac{\D[2]{y}}{\D{\zeta^2}}-Ry,
\end{equation}
and
\begin{align}
P& \equiv \frac{1}{\tilde{F}}\left(M\omega_0^2-\lambda_1\right)\left(z_2-z_1\right)^2,  \\ \qquad R &\equiv \frac{1}{\tilde{F}}Q_1 \omega_0^2 (z_2-z_1)^4 >0, \\
S & \equiv \frac{Q_1 \omega_0}{\tilde{F}}(z_2-z_1)^4(\omega_0 p_{20}+2\omega_1 p_{10})-\frac{1}{\tilde{F}}\left(2 \omega_0\omega_1M\right)\left(z_2-z_1\right)^2\frac{\D[2]{p_{10}}}{\D{\zeta^2}}, \label{definition_S}
\end{align}
subject to:
\begin{align}
\frac{\D{p_{11}}}{\D{\zeta}}&=0 \;\;\; \text{at} \;\;\; \zeta =0, \label{p_11_bc1_zeta}\\
\frac{z_1}{z_2-z_1}\frac{\D{p_{11}}}{\D{\zeta}}+p_{11}&=0 \;\;\; \text{at} \;\;\; \zeta =1,  \label{p_11_mixed_correction}\\
\frac{\D[2]{p_{11}}}{\D{\zeta^2}}&=0 \;\;\; \text{at} \;\;\; \zeta =0,1. \label{p_11_bc3_zeta}
\end{align}
At $O(\epsilon)$, the normalisation condition (\ref{normlisation_zeta}) becomes
\begin{equation}\label{p_11_normalisation}
\frac{z_1}{z_2-z_1} p_{10}'(1)\Bigl(p_{20}'(1)+p_{11}'(1)\Bigr)+\int_0^1 \frac{\D{p_{10}}}{\D{\zeta}}\left(\frac{\D{p_{20}}}{\D{\zeta}}+\frac{\D{p_{11}}}{\D{\zeta}}\right)\D{\zeta}=0.
\end{equation}

The operator $\mathcal{L}$ here is the same as in equation (\ref{p_10_scaled_governing}),  and $\omega_0$ is set so that the associated homogeneous problem permits non-trivial solutions. By the \textit{Fredholm Alternative} \citep{kress1989linear}, this means that a solution of (\ref{p_11_governing})--(\ref{p_11_normalisation}) will exist only when the solutions of the adjoint of the associated homogeneous problem to (\ref{p_11_governing})--(\ref{p_11_bc3_zeta}) are orthogonal to the inhomogeneous part $S$ of (\ref{p_11_governing}). This condition is known as a secularity condition. In this case, it sets the oscillation frequency correction, $\omega_1$,  as this is the only undetermined part of $S$. 

\subsubsection{The adjoint problem}
The associated homogeneous problem of (\ref{p_11_governing})--(\ref{p_11_normalisation}) is given by 
\begin{equation}\label{associated homogeneous}
\mathscr{L}(p_{11})=0,
\end{equation}
subject to the boundary conditions (\ref{p_11_bc1_zeta})--(\ref{p_11_normalisation}).

We define the inner product
\begin{equation}\label{standard_inner_product}
\langle u , v  \rangle =\int_0^1 u v \D{\zeta}.
\end{equation}
Using integration by parts three times and applying the boundary conditions (\ref{p_11_bc1_zeta})--(\ref{p_11_bc3_zeta}) on $p_{11}$, we can show that
\begin{align}\label{boundary_terms}
\langle \mathscr{L}p_{11}, v \rangle &= \langle p_{11}, \mathscr{L} v \rangle +\bigg[\frac{\D[3]{p_{11}}}{\D{\zeta^3}}v \bigg]_0^1+\bigg(P\frac{\D[2]{v}}{\D{\zeta^2}}+\frac{\D[3]{v}}{\D{\zeta^3}}\bigg)p_{11}\bigg|_{\zeta=0} \nonumber \\ 
&  \qquad \qquad \qquad \qquad +\bigg[\frac{\D[2]{v}}{\D{\zeta^2}}+\frac{z_1}{z_2-z_1}\bigg(P\frac{\D{v}}{\D{\zeta}}+ \frac{\D[3]{v}}{\D{\zeta^3}}\bigg)\bigg]\frac{\D{p_{11}}}{\D{\zeta}}\bigg|_{\zeta=1},
\end{align}
for any sufficiently differentiable function $v$. In order to derive the adjoint problem, we require the boundary terms present in (\ref{boundary_terms}) to vanish. This requirement sets the adjoint boundary conditions. 

We find that the adjoint problem for $v$ is then given by
\begin{equation}
\mathscr{L}(v)=0,
\end{equation}
subject to:
\begin{align}
v &= 0 \;\;\; \text{at} \;\;\; \zeta = 0,1, \label{adjoint_1} \\
P\frac{\D{v}}{\D{\zeta}}+\frac{\D[3]{v}}{\D{\zeta^3}}&=0 \;\;\; \text{at} \;\;\; \zeta =0, \\
\frac{\D[2]{v}}{\D{\zeta^2}}+\frac{z_1}{z_2-z_1}\bigg(P\frac{\D{v}}{\D{\zeta}}+\frac{\D[3]{v}}{\D{\zeta^3}}\bigg)& =0 \;\;\; \text{at} \;\;\; \zeta=1. \label{adjoint_3}
\end{align}

The solution for $v$ is sought in the same way as for $p_{10}$ (see \S\ref{sol_p_10_section}) and so we omit the detailed calculations here. We find that the solution is given by 
\begin{align}\label{eq:adj_sol}
v(\zeta) = A^\dag \frac{\sinh(g\zeta)}{\sinh(g)}+B^\dag \frac{\sinh(g(1-\zeta))}{\sinh(g)}+C^\dag\cos(h\zeta)+D^\dag\sin(h \zeta),
\end{align}
where:
\begin{align}
A^\dag &= \left(\frac{g\cosh(g)\sin(h)\left(P+g^2\right)+h\sinh(g)\cos(h)\left(P-h^2\right)}{h \sinh(g)\left(P-h^2\right)+g^3\sin(h)}\right)B^\dag, \\
C^\dag &=  -B^\dag, \\
D^\dag &= \left(\frac{h\sinh(g)\cot(h)\left(h^2-P\right)-g\cosh(g)\left(P+g^2\right)}{h \sinh(g)\left(P-h^2\right)+g^3\sin(h)}-\cot(h)\right)B^\dag.
\end{align}
The final constant $B^\dagger$ sets the amplitude of the solution. This is arbitrary here, so we can simply choose $B^\dagger=1$ for simplicity.
\subsubsection{Expression for $\omega_1$}
The secularity condition is $\langle S,v \rangle = 0$,  where $S$ is defined by (\ref{definition_S}) and $v$ is the adjoint solution (\ref{eq:adj_sol}).  Re-arranging the secularity condition for $\omega_1$, we find that this implies
\begin{align}\label{correction_oscillation_frequency}
\omega_1 = \frac{\left(z_2-z_1\right)^2 \omega_0 Q_1 \left\langle p_{20} , v \right\rangle}{2\left(M \left\langle \frac{\D[2]{p_{10}}}{\D{\zeta^2}}, v \right\rangle -Q_1\left(z_2-z_1\right)^2\left\langle p_{10}, v \right\rangle \right)}.
\end{align}

\begin{figure}
\centering
\includegraphics[width = \textwidth]{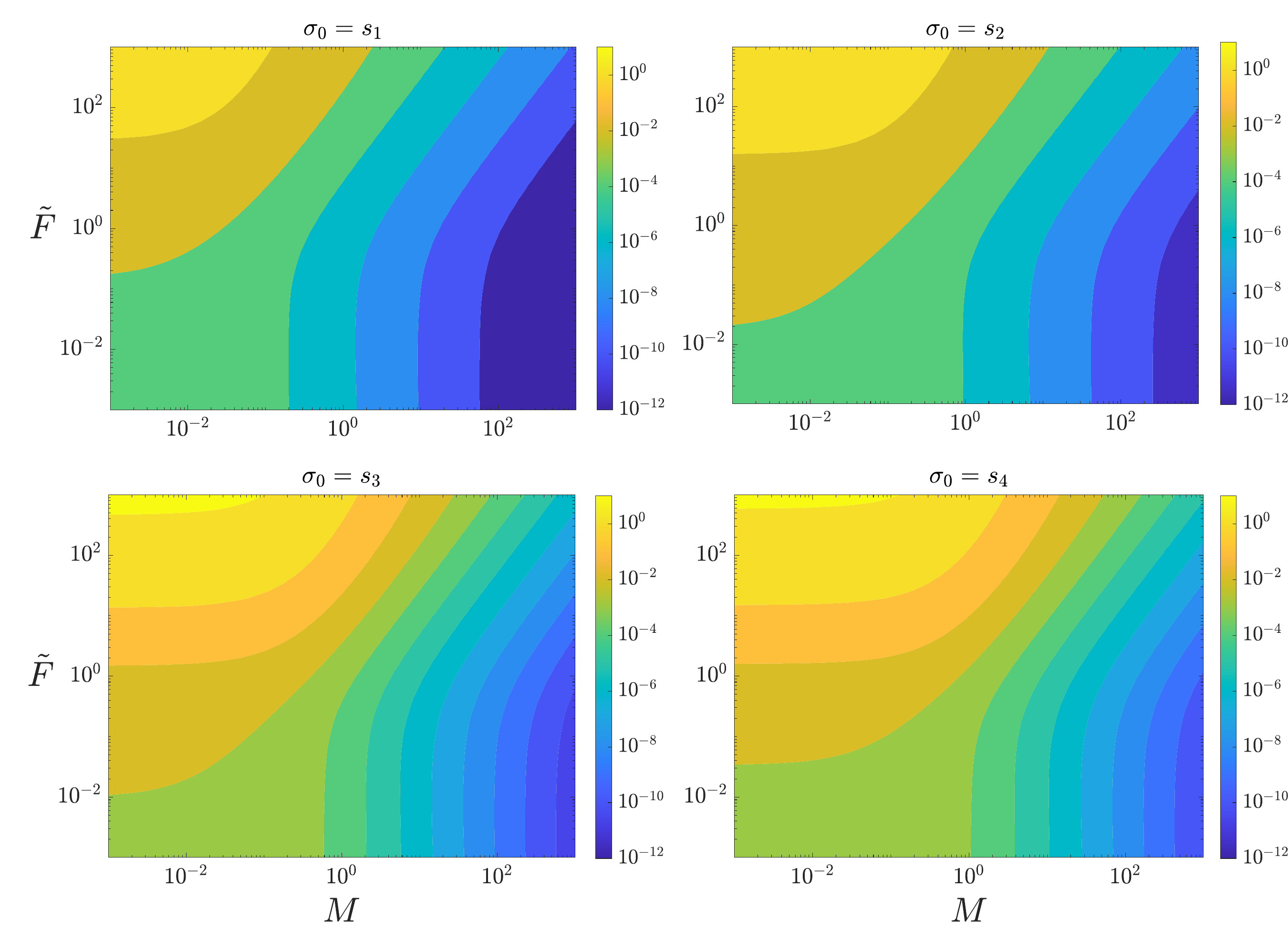}
\caption[The values $\epsilon\omega_1$, plotted throughout $(M, \tilde{F})$ parameter space.]{The values $\epsilon\omega_1$, plotted throughout $(M, \tilde{F})$ parameter space, for $\sigma_0 = s_1, s_2, s_3, s_4$. The values were calculated using the expression (\ref{correction_oscillation_frequency}) with $z_1 = 0.1, z_2 = 0.9$ and the values for $Q_1$ were obtained from the numerical data for $q_nt_n$,  which can be found in \cite{Netherwood2023deformations}}\label{omega_1_MF}
\end{figure}

In table \ref{correction-oscillation_table} we tabulate our results for $\epsilon \omega_1$ for a range of values of $M$, $\tilde{F}$ and $\sigma_0$.  In figure \ref{omega_1_MF},  we plot the solutions (\ref{correction_oscillation_frequency}) for $\epsilon\omega_1$. The plots maintain a similar profile to the leading-order component of the oscillation frequency, $\omega_0$, but with significantly smaller magnitudes observed throughout parameter space. Much like in figure \ref{omega_0_contour_plot}, an increase in $M$ results in a lower frequency, however the solutions for the correction $|\epsilon\omega_1|$ decay much faster and to smaller values. Examining the effect of different initial ellipticities, we again retain the feature observed in figure \ref{omega_0_contour_plot}, where we see an increase in the frequency with a decrease in $\sigma_0$. 
\begin{table}
\centering
\begin{tabular}{c|c|c|c|c|c}
\hline 
               & $\tilde{F}$\textbackslash{}$M$ & 0        & 0.1       & 1          & 10         \\ \hline 
 & 0.1                & $-0.00863$ & $-0.000441$ & $-2.88 \times 10^{-6}$ & $-9.84\times 10^{-9}$ \\ 
        $\sigma_0=s_1$       & 1                  & $-0.0345$  & $-0.00169$  & $-1.14\times 10^{-5}$ & $-3.93\times 10^{-8}$ \\ 
               & 10                 & $-0.389$   & $-0.0254$   & $-0.000199$  & $-7.02\times 10^{-7}$ \\ 
               & 100                & $-2.35$   & $-0.283$    & $-0.004584$  & $-1.98\times 10^{-5}$ \\ \hline
 & 0.1                & $-0.01367$ & $-0.00427$  & $-9.96\times 10^{-5}$ & $-4.22 \times 10^{-7}$ \\ 
         $\sigma_0=s_2$      & 1                  & $-0.0578$  & $-0.0176$   & $-0.000411$  & $-1.786\times 10^{-6}$ \\ 
               & 10                 &$ -0.656$   & $-0.231$    & $-0.00694 $  & $-3.27\times 10^{-5}$ \\ 
               & 100                & $-3.872$   & $-1.716$    & $-0.108$     & $-0.000861$  \\ \hline
& 0.1                & $ -0.0148$  & $-0.00814$  & $-0.000530$  & $-3.092 \times 10^{-6}$ \\ 
         $\sigma_0=s_3$       & 1                  & $-0.0666$  & $-0.0361$   & $-0.00233$   & $-1.403 \times 10^{-5}$ \\ 
               & 10                 &$ -0.756$   &$ -0.447$    &$ -0.0374$    & $-0.000263$  \\ 
               & 100                & $-4.306$   & $-2.846$    &$ -0.417$     & $-0.00617$   \\ \hline
& 0.1                & $-0.0131$  & $-0.00951$  &$ -0.00148$   &$ -1.40\times 10^{-5}$ \\ 
      $\sigma_0=s_4$          & 1                  & $-0.0619$  & $-0.0447$   & $-0.00685$   & $-6.72 \times 10^{-5}$ \\ 
               & 10                 &$ -0.697$   & $-0.529$    & $-0.101$     & $-0.00126$   \\ 
               & 100                & $-3.83$   & $-3.08$    & $-0.853$     & $-0.0250$     \\
\end{tabular} \caption[The values $\epsilon\omega_1$, tabulated for a variety of $M, \tilde{F}$ and $\sigma_0$.]{The values $\epsilon\omega_1$, tabulated for a variety of $M, \tilde{F}$ and $\sigma_0$. The values were obtained by substituting the solutions (\ref{omega_0_formula}), (\ref{general_solution_p_10}) and (\ref{p_20_solution}) for $\omega_0, p_{10}$ and $p_{20}$  into the analytical expression (\ref{correction_oscillation_frequency}) for $\omega_1$, with $z_1 = 0.1$ and $z_2=0.9$. The values for $Q_1=q_1t_1$ were obtained from the numerical data for $q_nt_n$,  which can be found in \cite{Netherwood2023deformations}.  }\label{correction-oscillation_table}
\end{table}
\subsubsection{Solution for $p_{11}$ }\label{p_11_solution_section}
Recall that $p_{11}$ is the solution of $\mathcal{L}(p_{11})=S$ as defined in (\ref{p_11_governing})--(\ref{definition_S}) subject to the boundary conditions (\ref{p_11_bc1_zeta})--(\ref{p_11_bc3_zeta}) and normalisation (\ref{p_11_normalisation}).  Note that $\mathcal{L}$ with these conditions is singular (i.e.,  permits non-trivial solutions to the homogeneous problem) due to the choice of $\omega_0$,  and that $S$ is in the image of $\mathcal{L}$ due to the choice of $\omega_1$.  

Since $\mathcal{L}(p_{10})=0$ and $p_{10}$ satisfies the boundary conditions (\ref{p_11_bc1_zeta})--(\ref{p_11_bc3_zeta}),  we can write the solution for $p_{11}$ as
\begin{equation}\label{solution_p_11}
p_{11}(\zeta)= \alpha p_{10}(\zeta)+p^\star(\zeta),
\end{equation}
where $p^\star$ is a particular solution of $\mathcal{L}(p_{11})=S$ subject to (\ref{p_11_bc1_zeta})--(\ref{p_11_bc3_zeta}),  and $\alpha$ is chosen to ensure that the normalisation condition (\ref{p_11_normalisation}) is satisfied.  

Obtaining a solution for $p^\star$ is complicated by the singular nature of the system,  and the fact that $\omega_0$ and $\omega_1$ have been determined numerically.  The numerical error in $\omega_0$ means that,  in practice,  the system is not quite singular.  The numerical error in $\omega_1$ means that the right hand side $S$ is not quite in the true image of $\mathcal{L}$.  These issues can combine to give significant errors in the solution for $p^\star$ obtained by solving this system.  

In order to avoid these difficulties,  we can instead consider an equivalent non-singular and well-posed problem for $p^\star$.  First,  the singular nature of the problem is removed by removing the mixed boundary condition (\ref{p_11_mixed_correction}).  Since $S$ is in the image of $\mathcal{L}$,  the solution obtained for $p^\star$ will still satisfy this final boundary condition.  Secondly,  we ensure a well-posed problem for $p^\star$ by adding a new homogeneous boundary condition at $\zeta =0$,  which is not satisfied by $p_{10}$. The revised problem for $p^\star$ is thus: 
\begin{align}\label{H_governing}
\mathscr{L}(p^\star)=\frac{1}{\tilde{F}}Q_1\omega_0(z_2-z_1)^4( \omega_0 p_{20}+2\omega_1 p_{10})-\frac{1}{\tilde{F}}\left(2 \omega_0\omega_1M\right)\left(z_2-z_1\right)^2\frac{\D[2]{p_{10}}}{\D{\zeta^2}},
\end{align}
subject to:
\begin{align}
p^\star&=0 \qquad  \text{at} \qquad \zeta =0, \label{H_bc1_zeta}\\
\frac{\D{p^\star}}{\D{\zeta}}&=0 \qquad \text{at} \qquad  \zeta =0, \\
\frac{\D[2]{p^\star}}{\D{\zeta^2}}&=0 \qquad \text{at} \qquad  \zeta =0,1. \label{H_bc3_zeta}
\end{align}
The formal solution to (\ref{H_governing})--(\ref{H_bc3_zeta}) still satisfies the original problem for $p^\star$,  however,  the practical solution is now much less sensitive to the small numerical errors in $\omega_0$ and $\omega_1$. 

The system (\ref{H_governing})--(\ref{H_bc3_zeta}) is solved analytically by writing the general solution for $p^\star$ in the form 
\begin{align}\label{p_11_star_general_sol}
p^\star(\zeta) &= A^\star \frac{\sinh(g \zeta)}{\sinh(g)}+B^\star \frac{\sinh(g(1- \zeta))}{\sinh(g)} +C^\star\cos (h \zeta) + D^\star \sin (h \zeta) \nonumber \\ & \qquad  + C_{1}^\star+C_{2}^\star\zeta + C_{3}^\star\frac{\sinh(\psi \zeta)}{\sinh(\psi)} + C_{4}^\star \frac{\sinh(\psi (1-\zeta))}{\sinh(\psi)} + C_{5}^\star \frac{\zeta \sinh(g\zeta)}{\sinh(g)} \nonumber \\ &  \qquad \qquad  + C_{6}^\star \frac{\zeta \sinh(g(1- \zeta))}{\sinh(g)} + C_{7}^\star \zeta \cos(h \zeta) + C_{8}^\star \zeta \sin(h \zeta),
\end{align}
where $A^\star, B^\star,C^\star,D^\star$ are arbitrary constants in the complimentary function and $C_1^\star, C_2^\star, \dots C_8^\star$ are constants present in the particular integral.  On substituting the particular integral into the governing equation (\ref{H_governing}) and using \texttt{Maple} to equate coefficients, we obtain analytical expressions for the constants $C_1^\star, C_2^\star, \dots C_8^\star$ in terms of $z_1,z_2, M, \tilde{F}$ and the numerically determined constants $h,g,\omega_0,Q_1$ and $Q_2$. 

With the particular integral known, we can apply the boundary conditions (\ref{H_bc1_zeta})--(\ref{H_bc3_zeta}) to the solution (\ref{p_11_star_general_sol}). Explicit expressions for the constants $A^\star, B^\star,C^\star$ and $D^\star$ can then be obtained using in terms of $z_1,z_2, M, \tilde{F}$ and the numerically determined constants $h,g,\omega_0,Q_1$ and $Q_2$ using \texttt{Maple}.

The pressure correction is given by (\ref{solution_p_11}). Having determined $p^\star$ uniquely,  the normalisation condition (\ref{p_11_normalisation}) then determines the remaining unknown, $\alpha$.  Substituting (\ref{solution_p_11}) into (\ref{p_11_normalisation}) and making use of (\ref{leading_order_normalisation}),  we find that
\begin{align}\label{p_11_normalisation_alpha}
\alpha & = -\frac{1}{z_1-z_2}\int_0^1 \frac{\D{p_{10}}}{\D{\zeta}}\left(\frac{\D{p_{20}}}{\D{\zeta}} +\frac{\D{p^\star}}{\D{\zeta}}\right)\D{\zeta}\nonumber \\ & \qquad \qquad -\frac{z_1}{(z_2-z_1)^2}\left[\frac{\D{p_{10}}}{\D{\zeta}}\left(\frac{\D{p_{20}}}{\D{\zeta}} +\frac{\D{p^\star}}{\D{\zeta}}\right)\right]\bigg|_{\zeta =1}.
\end{align}
We choose not to present the full analytical solution for $p_{11}$ due to the expressions for the coefficients being prohibitively complex.

\begin{figure}
\centering 
\includegraphics[scale=0.23]{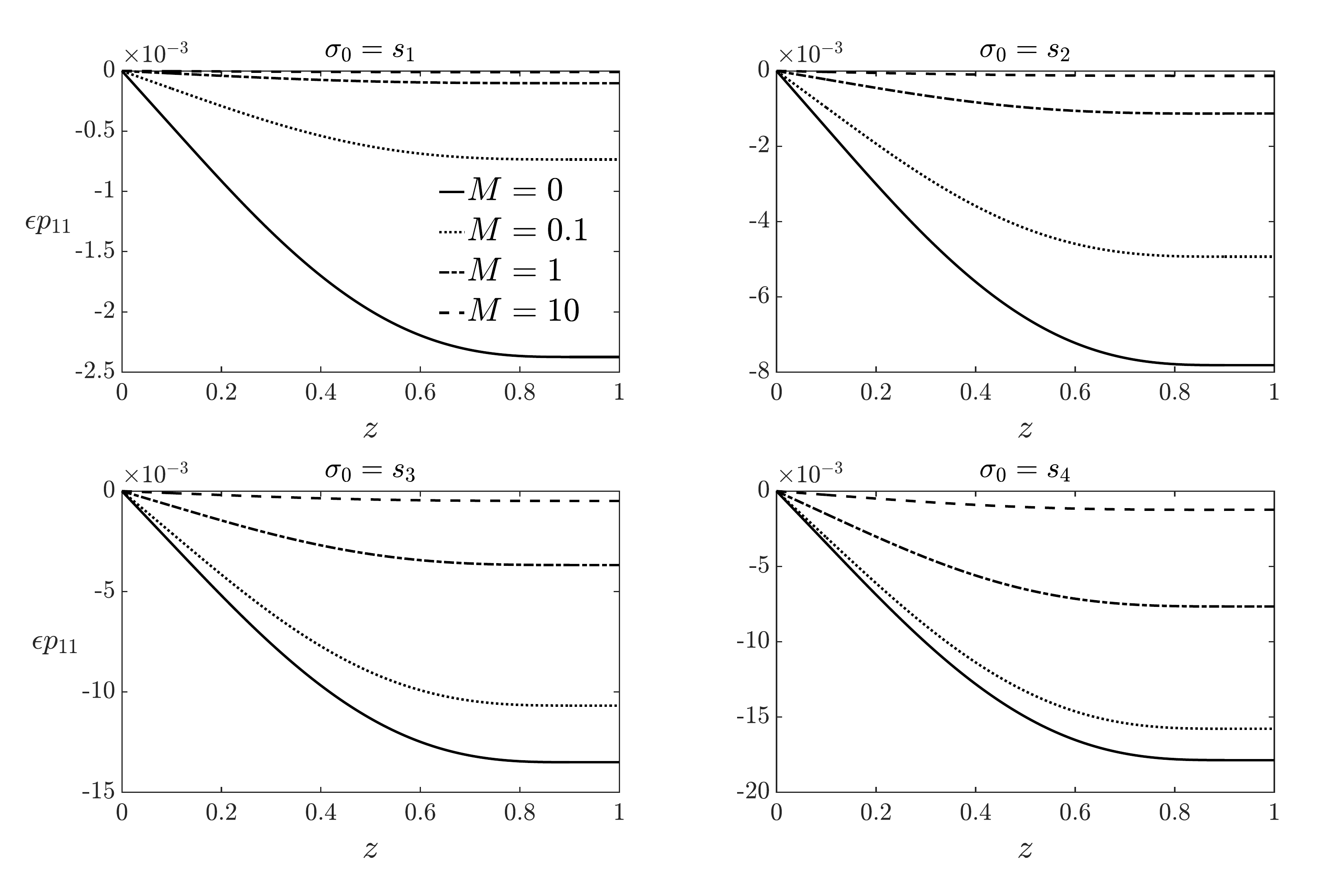}
\caption[Solutions for $\epsilon p_{11}$.]{Solutions for $\epsilon p_{11}$, plotted for $\tilde{F}=1$, $z_1=0.1,z_2=0.9$ and varying values of $M$ and $\sigma_0$.  The plots were obtained using the analytical expressions (\ref{solution_p_11}) and (\ref{p_11_star_general_sol}), with analytical solutions for the coefficients and $\alpha$ determined using \texttt{Maple}.  The values of the representative $\sigma_0 = s_i$ are given in table \ref{Q_epsilon_table_values}.  Note the different scales on the vertical axes.}\label{p_11_figure}
\end{figure}

\begin{figure}
\centering
\includegraphics[width=\textwidth]{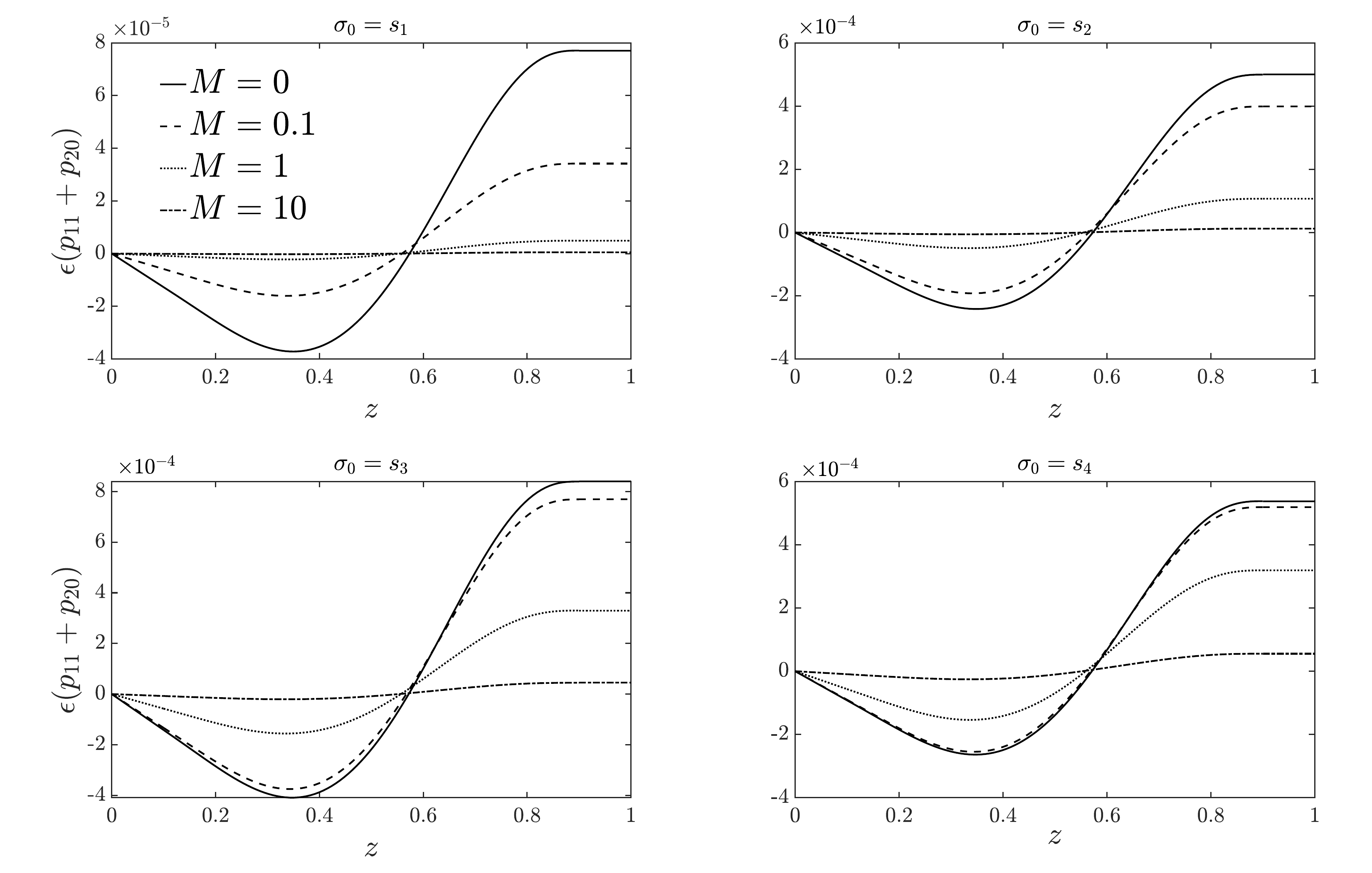}
\caption[Solutions for $\epsilon(p_{20}+ p_{11})$.]{Solutions for $\epsilon(p_{20}+ p_{11})$, plotted for $\tilde{F}=1$, $z_1=0.1,z_2=0.9$ and varying values of $M$ and $\sigma_0$. The solution for $p_{20}$ was obtained using (\ref{p_20_solution})--(\ref{C_4}), and for $p_{11}$ we use (\ref{solution_p_11}) together with (\ref{p_11_star_general_sol}). Each of the constants involved are known in terms of $z_1,z_2, M, \tilde{F}$ and the numerically determined constants $h,g,\omega_0,Q_1$ and $Q_2$.  }\label{p_11_plus_p20_plot}
\end{figure}

Solutions for $\epsilon p_{11}$ are plotted in figure \ref{p_11_figure}. The features of the plots are similar to those observed in figure \ref{p_20_solution_figure}. We find that changes in the inertia coefficient amount to an observable difference in the profiles $\epsilon p_{11}$.  The solutions with maximal amplitude correspond to $M=0$. In figure \ref{p_11_plus_p20_plot},  we plot the full $O(\epsilon)$ component $\epsilon(p_{20}+p_{11})$ against $z$ for $\sigma_0 = s_1,s_2,s_3,s_4$.  The plots demonstrate that varying both the inertia coefficient and ellipticity parameter have significant impact on the amplitude of the correction.
\subsection{Truncation error estimates}\label{error_analysis_section}
\begin{figure}
\centering
\includegraphics[width=\textwidth]{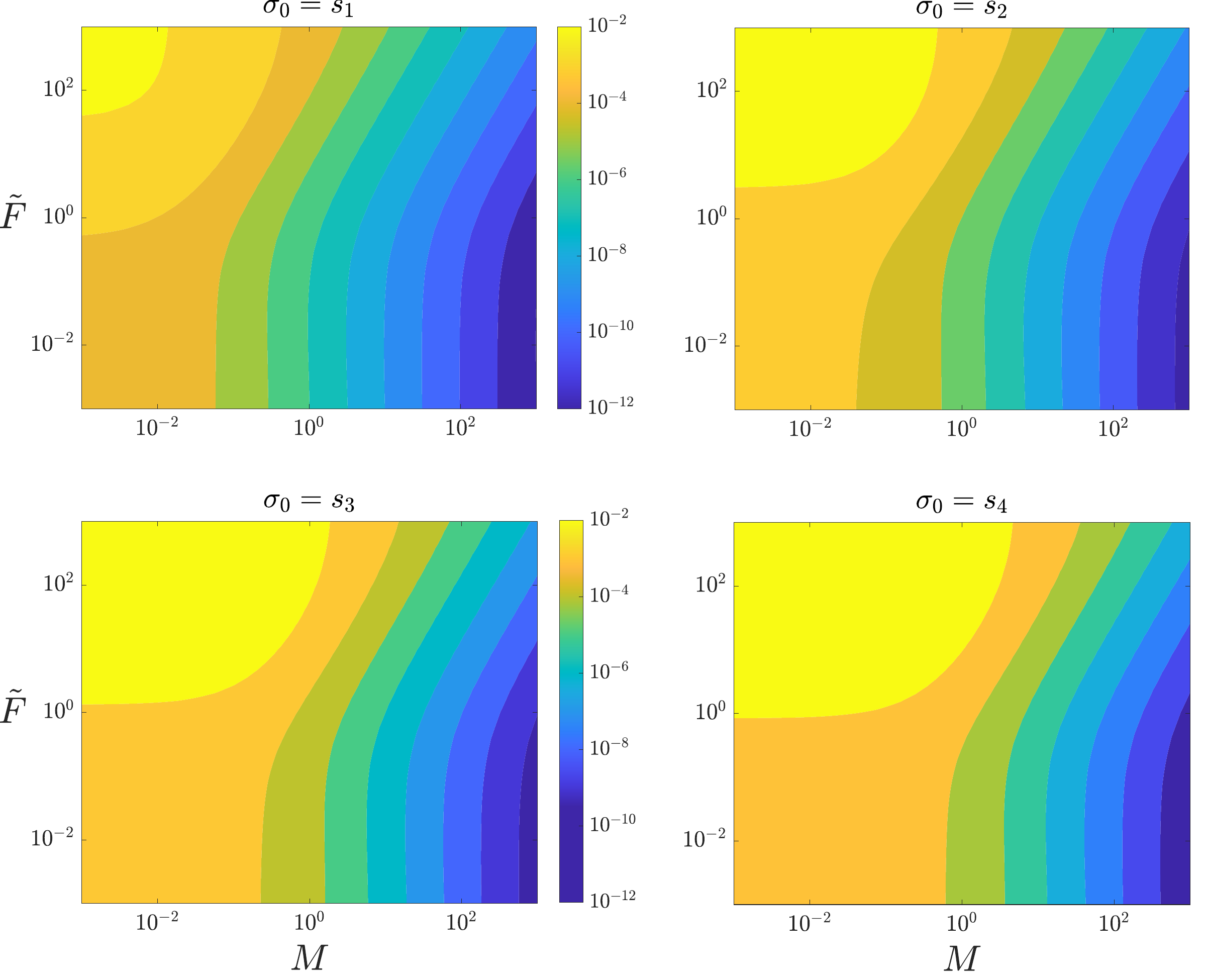}
\caption[The values $|\epsilon \omega_1/\omega_0|$,  plotted throughout ($M, \tilde{F})$ parameter space.]{The values $|\epsilon \omega_1/\omega_0|$, plotted throughout regions of ($M, \tilde{F})$ parameter space. The values were computed using the analytical expressions (\ref{omega_0_formula}) and (\ref{correction_oscillation_frequency}). The results illustrate the error after truncating (\ref{full_omega}) after $O(\epsilon^0)$.}\label{Frequency_error_MF}
\end{figure}

\begin{figure}
\centering
\includegraphics[width = \textwidth]{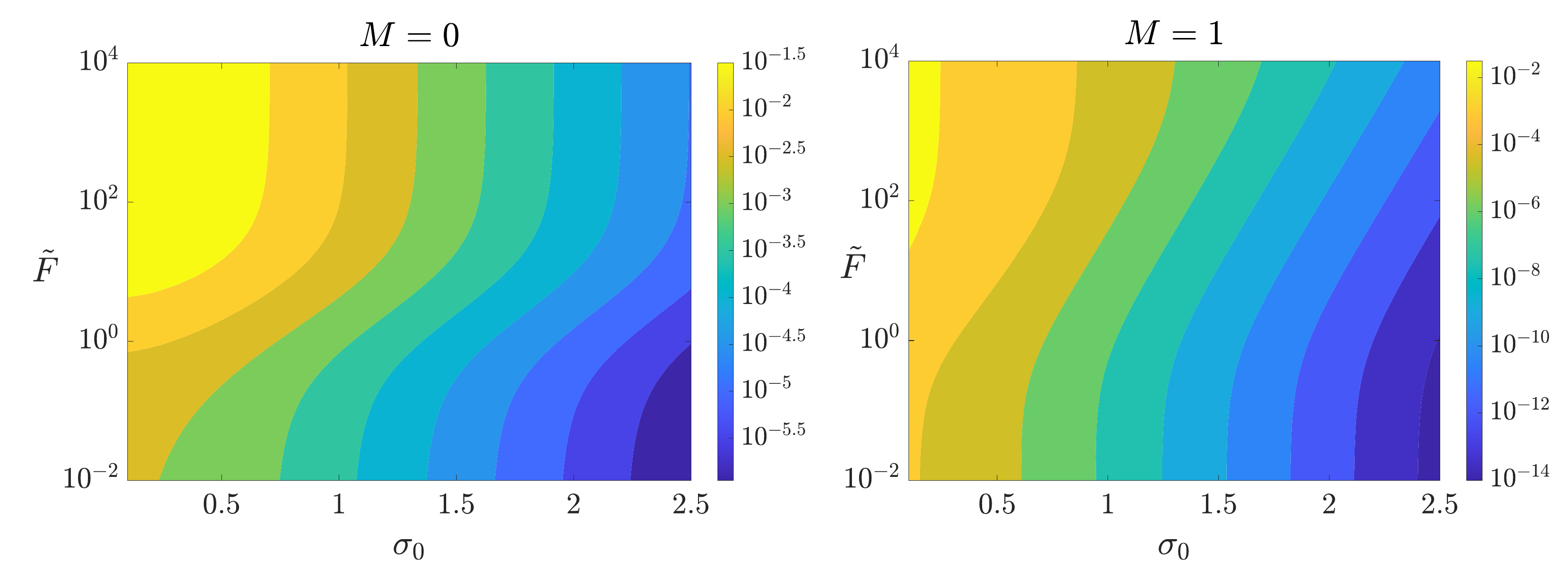}
\caption[The values $|\epsilon \omega_1/\omega_0|$, plotted throughout ($\tilde{F},\sigma_0)$ parameter space.]{The values $|\epsilon \omega_1/\omega_0|$, plotted throughout regions of ($\tilde{F},\sigma_0)$ parameter space, with $M=0,1$. The values were computed using the analytical expressions (\ref{omega_0_formula}) and (\ref{correction_oscillation_frequency}). The results illustrate the error after truncating (\ref{full_omega}) after $O(\epsilon^0)$. }\label{Frequency_error_Fsigma}
\end{figure}

With solutions for $p_{10}, p_{20}, p_{11}, \omega_0$ and $\omega_1$ known for general $\sigma_0, M$ and $\tilde{F}$, we can now investigate the error incurred by truncating the series expansions:
\begin{align}\label{full_p}
\tilde{p}(z) &=p_{10}+\epsilon\left(p_{11}+p_{20} \right)+ O(\epsilon^2), \\
\omega &=\omega_0+\epsilon \omega_1+ O(\epsilon^2),\label{full_omega}
\end{align}
after $O(\epsilon^0)$.
In figures \ref{Frequency_error_MF} and \ref{Frequency_error_Fsigma} we plot the values $|\epsilon \omega_1/\omega_0|$ in $(M,\tilde{F})$ and $(\tilde{F},\sigma_0)$ parameter space respectively, in order to gain an understanding of the errors incurred by truncating the expansion (\ref{full_omega}) for the oscillation frequency after $O(\epsilon^0)$. The results show that changes in $M,\tilde{F}$ and $\sigma_0$ have a significant impact on the magnitude of the relative error. 

Figure \ref{Frequency_error_MF} shows that an increase in $M$ results in a decrease in the relative error, whilst an increase in $\tilde{F}$ amounts to an increase in the relative error. Examining variations in $\sigma_0$ in figure \ref{Frequency_error_Fsigma}, we see that for fixed $M$, the error decreases monotonically with increasing $\sigma_0$ and/or decreasing $\tilde{F}$. As $\sigma_0 \to \infty$ (i.e., as the tube's initial cross section becomes circular) the relative error approaches zero. In the large and small $\tilde{F}$ limits, we observe that the relative error becomes independent of $\tilde{F}$.  

\begin{figure}
\centering 
\includegraphics[width=\textwidth]{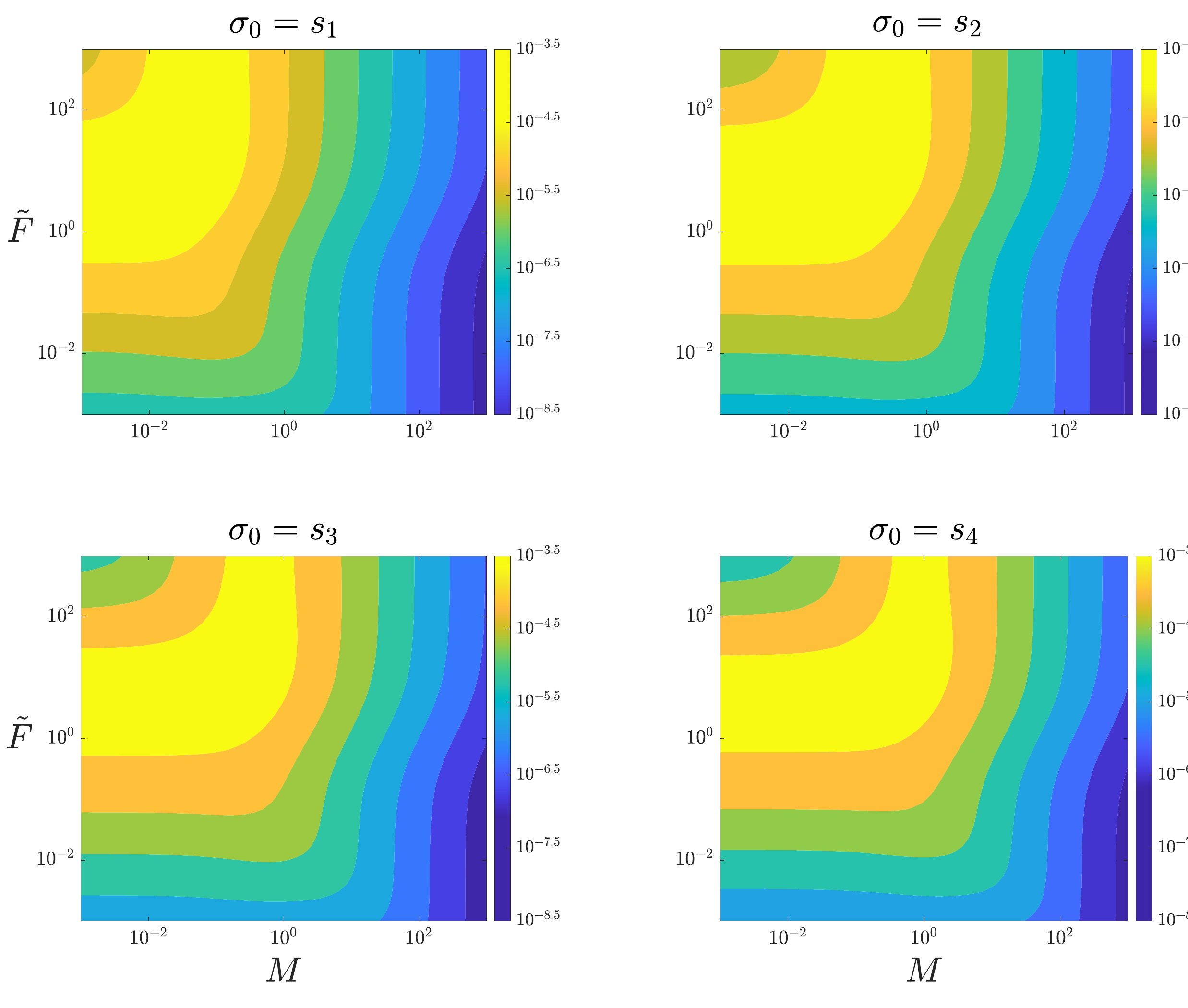}
\caption[The values $|\epsilon\left(p_{11}+p_{20} \right)/p_{10}|$ plotted throughout $(M,\tilde{F})$ parameter space.]{The values $|\epsilon\left(p_{11}+p_{20} \right)/p_{10}|$ evaluated at $z=0.7$, plotted throughout regions of $(M,\tilde{F})$ parameter space. The values were computed using the analytical expressions (\ref{general_solution_p_10}), (\ref{p_20_solution}) and (\ref{solution_p_11}) for $p_{10}$, $p_{11}$ and $p_{20}$ respectively. The results give an indication of the error incurred by truncating (\ref{full_p}) after $O(\epsilon^0)$.}\label{Total_error_MF}
\end{figure}

In figure \ref{Total_error_MF} we plot the values $\epsilon\left(p_{11}+p_{20} \right)/p_{10}$, evaluated at $z=0.7$, throughout $(M,\tilde{F})$ parameter space. The results give an understanding of the error incurred by truncating the expansion (\ref{full_p}) for $\tilde{p}$ after $O(\epsilon^0)$. Much like for the frequency, our results show that larger values of the inertia coefficient, $M$, result in a smaller relative error.  Consistent with \cite{Netherwood2023deformations}, we find that smaller values of $\sigma_0$, which correspond to more non-circular initial cross-sectional shapes, result in larger magnitudes of the relative error.  Examining variations in the axial tension, we see that for fixed $M \lesssim O(1)$ the relative error is maximal at around $\tilde{F} = O(1)$ with $M = 0$, and decays in the large and small $\tilde{F}$ limits. 

\begin{figure}
\centering 
\includegraphics[scale=0.24]{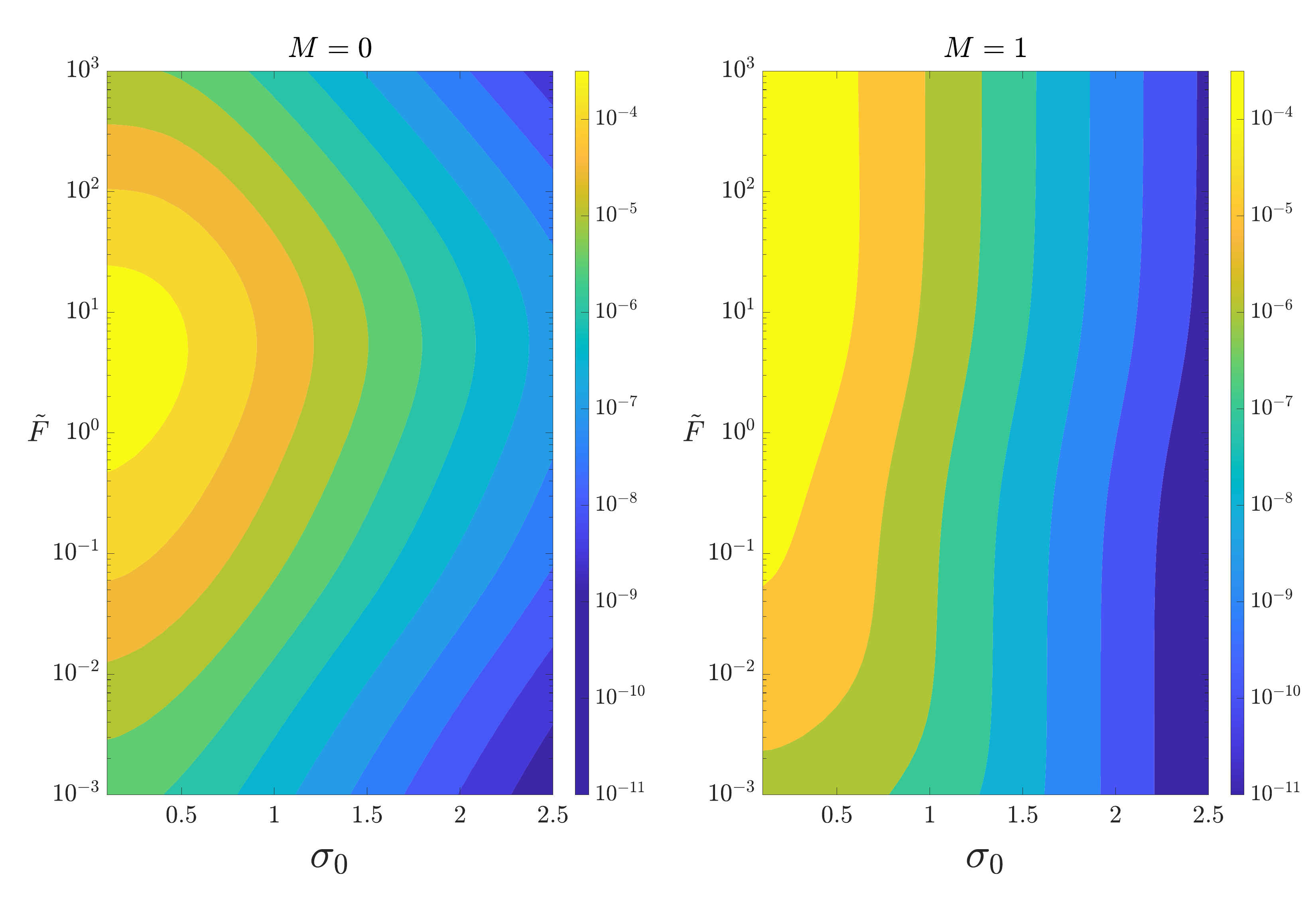}
\caption[The values $|\epsilon\left(p_{11}+p_{20} \right)/p_{10}|$ plotted throughout $(\tilde{F},\sigma_0)$ parameter space. ]{The values $|\epsilon\left(p_{11}+p_{20} \right)/p_{10}|$ evaluated at $z=0.7$, throughout regions of $(\tilde{F},\sigma_0)$ parameter space, with $M=0$ (top) and $M=1$ (bottom). The values were computed using the analytical expressions (\ref{general_solution_p_10}), (\ref{p_20_solution}) and (\ref{solution_p_11}) for $p_{10}$, $p_{11}$ and $p_{20}$. The results illustrate the error after truncating (\ref{full_p}) after $O(\epsilon^0)$.}\label{Total_error_F_sigma}
\end{figure}

In figure \ref{Total_error_F_sigma} we again plot values of $|\epsilon\left(p_{11}+p_{20} \right)/p_{10}|$, evaluated at $z=0.7$, however this time we fix $M$ whilst varying $\tilde{F}$ and $\sigma_0$. The results show very different behaviour of the error, depending on whether $M =0$ or $M \neq 0$. When $M=0$, we see that the error will decrease for both large and small $\tilde{F}$. However, when $M=1$, the error only decreases as $\tilde{F} \to 0$, and tends towards being independent of $\tilde{F}$ as $\tilde{F}\to \infty$.  In either case, we see that an increase in $\sigma_0$ leads to a smaller relative error. 
\section{Stability threshold and growth rate for self-excited oscillations}\label{stability_section}
The normal-mode solutions obtained in \S4 represent the solution
to the oscillatory problem at leading order in the dimensionless
amplitude $\delta$, aspect ratio $\ell^{-1}$, inverse Strouhal
number $St^{-1}$, and inverse Womersley number $\alpha^{-1}$.
These solutions oscillate sinusoidally and have constant
amplitude in time.   Higher-order corrections will lead to more
complicated time-dependence,  and changes in the amplitude over
longer time-scales.  However,  \cite{whittaker2011energetics} showed that the growth/decay rates of the normal-mode solutions can be determined by considering the global energy budget of the system, without having to compute any higher-order corrections explicitly.  For
a given set of problem parameters,  if there is a normal mode that
grows in time,  then the system will be unstable to this mode. If all the modes decay,  we expect the system to be stable.  

In this section, we show how the energetic approach of \cite{whittaker2011energetics} can be applied to the
current formulation. We will find an expression for the growth
rate of each mode, and a stability criterion in the form of a
critical Reynolds number for the steady mean flow through the
tube.
\subsection{Energy fluxes and fluid energy}
We adopt the results of \citet{whittaker2010predicting},  who obtained an expression for the tube's energy budget,  which has been averaged over one oscillation period.  The energy budget of the oscillatory perturbation was found to be
\begin{equation}\label{energy_budget}
\diff{}{t}\left(\tilde{\mathbb{E}}_s+\tilde{\mathbb{E}}_f \right) = \frac{1}{\ell St} \left(\mathscr{T} - \mathscr{P} - \mathscr{D}\right).
\end{equation}
Here $\tilde{\mathbb{E}}_s$ is the dimensionless energy due to oscillations of the tube wall and $\tilde{\mathbb{E}}_f$ is the dimensionless oscillatory kinetic energy in the fluid,  both of which have been time-averaged over a single oscillation period.  The right-hand side of (\ref{energy_budget}) features $\mathscr{T}$ as the dimensionless mean flux of kinetic energy through the tube ends due to the oscillatory flow,  $\mathscr{P}$ as the dimensionless mean rate of working by the pressure at the tube ends due to the oscillatory flow,  and $\mathscr{D}$ as the dimensionless mean rate of dissipation by the oscillatory flow.  Energy fluxes due to the mean flow were found to cancel out.  

\citet{whittaker2010predicting} obtained expressions for $\mathscr{T}$,  $\mathscr{P},\mathscr{D}$ and $\tilde{\mathbb{E}}_f$ initially in terms of the oscillatory axial fluid velocity,  before re-arranging (using (\ref{oscialltory_inertial_balance})) to obtain equivalent expressions in terms of the oscillatory pressure.  The explicit expressions,  which are still valid here,  were found to be:
\begin{align}
\mathscr{T} &= \frac{3}{4 \omega^2} \pi \ell^2 St^2 \Delta^2| \tilde{p}'(0)|^2, \label{kinetic_influx} \\
\mathscr{P} &= \frac{1}{4} \pi \ell^2 St^2 \Delta^2| \tilde{p}'(0)|^2,  \\
\mathscr{D} &= \frac{\pi \ell^3 St^3 \Delta^2 (2\omega)^{1/2}}{2 \alpha \omega^2}\int_0^1 |\tilde{p}'(z)|^2 \D{z}, \label{Dissipation_energy} \\
\tilde{\mathbb{E}}_f &= \frac{\Delta^2 St^2 A_0 \ell^2}{4 \omega^2}\int_0^1 |\tilde{p}'(z)|^2 \D{z}, \label{oscillatory_fluid_energy}
\end{align}
where $\tilde{p}(z)$ is the axial mode for the fluid pressure,  and is defined in terms of the solutions found in \S\ref{unsteady_coupled_problem_section} through equation (\ref{pressure_expansion}).

\citet{whittaker2011energetics} showed that the sufficiently large timescale on which the oscillations will grow/decay results in only the leading-order oscillatory fluid flow being required to calculate the expressions (\ref{kinetic_influx})--(\ref{oscillatory_fluid_energy}).  They argue that this means that slow changes in the steady flow do not contribute to the energy budget (\ref{energy_budget}) at leading order.  We therefore inherit the expressions (\ref{kinetic_influx})--(\ref{oscillatory_fluid_energy}) in the current work.  

To make use of (\ref{energy_budget}),  we need to determine an expression for $\tilde{\mathbb{E}}_s$,  the mean energy in the wall due to oscillations.  The procedure set out by \cite{whittaker2010predicting} and \cite{walters2018effect} to obtain $\tilde{\mathbb{E}}_s$ cannot be identically replicated when higher-order azimuthal modes are retained,  as is the case in the present work,  due to coupling between the area-change components,  $A_n$.  We therefore present an alternative approach in Appendix \ref{sec:energy_appendix},  where it is found that 
\begin{equation}\label{E_s_main_body}
\tilde{\mathbb{E}}_{s} = \frac{\Delta^2 St^2 \ell^2A_0}{4 \omega^2}\int_{0}^{1} \bigg|\sum_{n=1}^\infty\tilde{p}_n'\bigg|^2 + \sum_{n=1}^\infty \frac{2M}{q_nt_n} |\tilde{p}_n''|^2\D{z}.
\end{equation}
For the case in which the higher-order azimuthal modes ($n >1$) are neglected,  equation (\ref{E_s_main_body}) has an identical form to that obtained by \cite{walters2018effect},  and \cite{whittaker2010predicting} (for the case $M=0$),  but differs in the value of the numerical constant $q_1t_1$.  
\subsection{Expression for the growth rate and critical Reynolds number}
Summing the expressions (\ref{E_s_main_body}) and (\ref{oscillatory_fluid_energy}) for $\tilde{\mathbb{E}}_s$ and $\tilde{\mathbb{E}}_f$,  we obtain
\begin{equation}\label{E_s_plus+E_f}
\tilde{\mathbb{E}}_f + \tilde{\mathbb{E}}_s = \frac{\Delta^2 St^2 \ell^2 A_0}{2 \omega^2} \int_0^1 \bigg|\sum_{k = 1}^\infty \tilde{p}_k' \bigg|^2 +\sum_{n =1}^\infty \frac{M}{q_nt_n} |\tilde{p}_n''|^2\D{z}.
\end{equation}
Substituting (\ref{kinetic_influx})--(\ref{Dissipation_energy}) and (\ref{E_s_plus+E_f}) into the energy budget (\ref{energy_budget}),  rearranging,  and then writing any terms involving the global pressure as a sum over the pressure modes $p_i$,  it can be shown that
\begin{equation}\label{growth_rate_equation}
\diff{\Delta}{t} = \frac{\pi}{2 A_0} \left(\frac{\frac{1}{\ell St}|\sum_{n = 1}^\infty \tilde{p}_n'(0)|^2 - \frac{(2\omega)^{\frac{1}{2}}}{\alpha} \int_0^1 |\sum_{n = 1}^\infty \tilde{p}_n'|^2 \D{z}}{\int_0^1 |\sum_{n = 1}^\infty\tilde{p}_n'|^2+ \sum_{n = 1}^\infty \frac{M}{q_nt_n}|\tilde{p}_n''|^2\D{z}} \right) \Delta.
\end{equation}
Equation (\ref{growth_rate_equation}) demonstrates that the amplitude of the oscillatory normal modes will grow or decay exponentially with a dimensionless growth rate defined as 
\begin{equation}\label{generalised_growth_rate}
\Lambda = \frac{\pi}{2 A_0} \left(\frac{\frac{1}{\ell St}|\sum_{n = 1}^\infty \tilde{p}_n'(0)|^2 - \frac{(2\omega)^{\frac{1}{2}}}{\alpha} \int_0^1 |\sum_{n = 1}^\infty \tilde{p}_n'|^2 \D{z}}{\int_0^1 |\sum_{n = 1}^\infty\tilde{p}_n'|^2+ \sum_{n = 1}^\infty \frac{M}{q_nt_n}|\tilde{p}_n''|^2\D{z}} \right).
\end{equation}
The expression (\ref{generalised_growth_rate}) generalises the growth rate expressions obtained by \cite{whittaker2010predicting} and \cite{walters2018effect} by including contributions from the higher-order azimuthal modes. 

Observing (\ref{energy_budget}),  we can see that the growth or decay of the oscillations will depend on the sign of $\Lambda$.  Setting $\Lambda=0$ and recalling $\alpha^2/St = Re$,  the expression for the critical Reynolds number $Re_c$ is given as
\begin{equation}\label{critical_reynolds_number}
Re_c = \frac{\alpha \ell (2\omega)^{1/2}}{|\tilde{p}'(0)|^2}\int_0^1 |\tilde{p}'(z)|^2 \D{z} = \frac{\alpha \ell (2\omega)^{1/2}}{|\sum_{n=1}^\infty p'_n(0)|^2}\int_0^1 \bigg|\sum_{n=1}^\infty p'_n(z)\bigg|^2 \D{z}.
\end{equation}
For $Re<Re_c$,  we have $\Lambda<0$ and hence the flow is stable.  For $Re>Re_c$ we have $\Lambda>0$ and hence the flow is unstable.  It is noted that the fundamental azimuthal mode provides the largest contribution to $\Lambda$ and hence determines the critical Reynolds number and the overall stability.  

\begin{figure}
\centering
\includegraphics[scale=0.2]{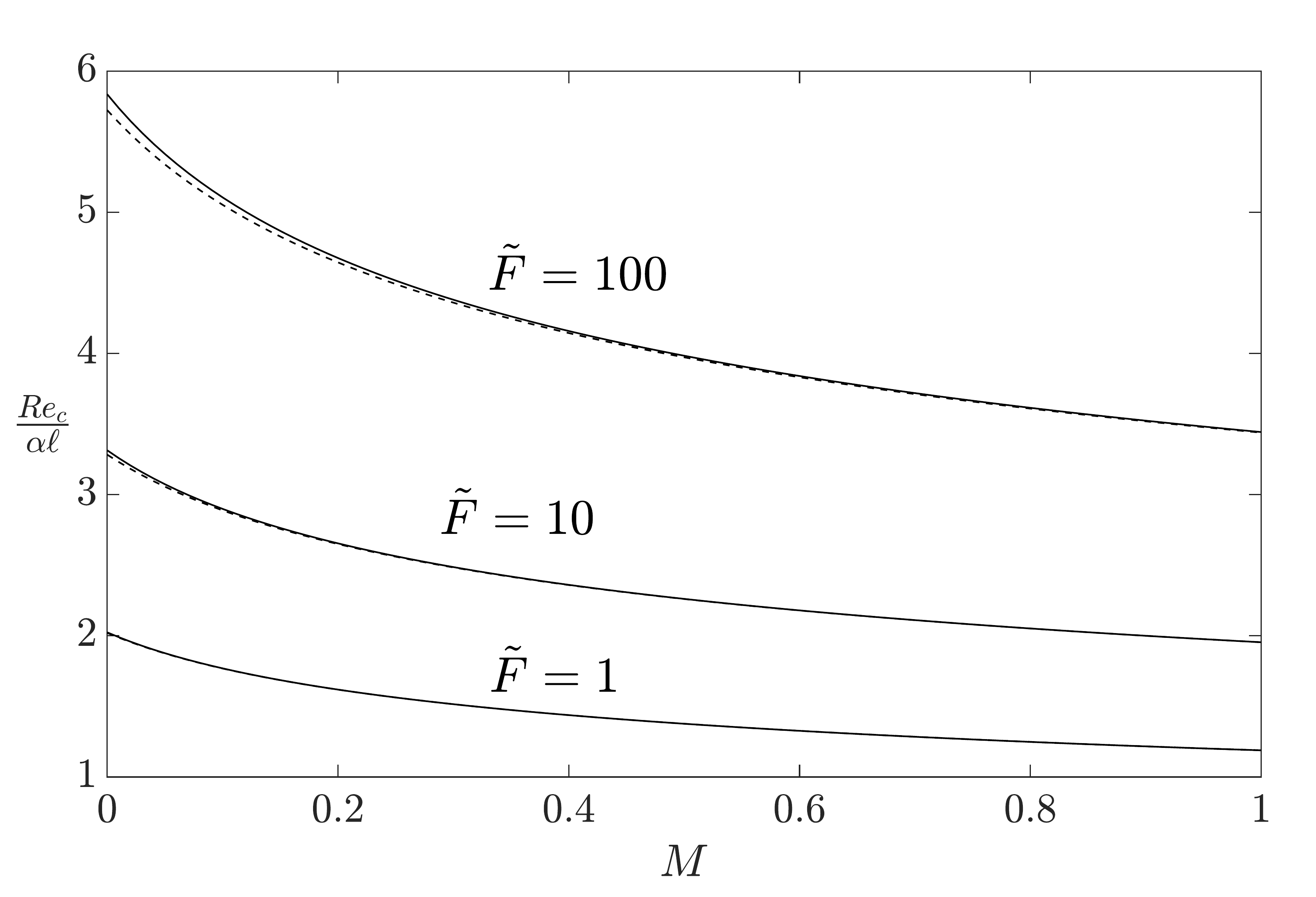}
\caption[The critical Reynolds number $Re_c/(\alpha \ell)$ plotted against $M$.]{The critical Reynolds number $Re_c/(\alpha \ell)$ plotted against $M$ with $z_1=0.1$,  $z_2 = 0.9$,  $\tilde{F}=1,10,100$,  and $\sigma_0 = 0.6$.  These solutions were obtained by substituting the expansions (\ref{full_p}) and (\ref{full_omega}) for the pressure and oscillation frequency truncated after $O(1)$ (dashed curves) and $O(\epsilon)$ (continuous curves) into the expression (\ref{critical_reynolds_number}) for the critical Reynolds number.  All of the solutions correspond to the fundamental oscillatory mode,  which is the most unstable. }\label{crit_reynolds_plot}
\end{figure}

In figure \ref{crit_reynolds_plot},  we plot solutions for the critical Reynolds number (\ref{critical_reynolds_number}) against the inertia coefficient $M$ by substituting the expansions (\ref{full_p}) and (\ref{full_omega}) for  $\tilde{p}$ and $\omega$ into (\ref{critical_reynolds_number}).  We include both the leading order and $O(\epsilon)$ approximations for $\tilde{p}$ and $\omega$ to give an indication of the impact the correction to the fundamental azimuthal mode has on the solution.  Our results are in good agreement with \citet{walters2018effect} (and \citet{whittaker2010predicting} for the case $M=0$),  which verifies the assertion that solutions corresponding to the fundamental azimuthal mode provide a good approximation for the critical Reynolds number.  Our results demonstrate that for physically realistic parameter values,  the contributions from the higher-order azimuthal modes can be neglected. 

\begin{figure}
\centering
\includegraphics[scale=0.2]{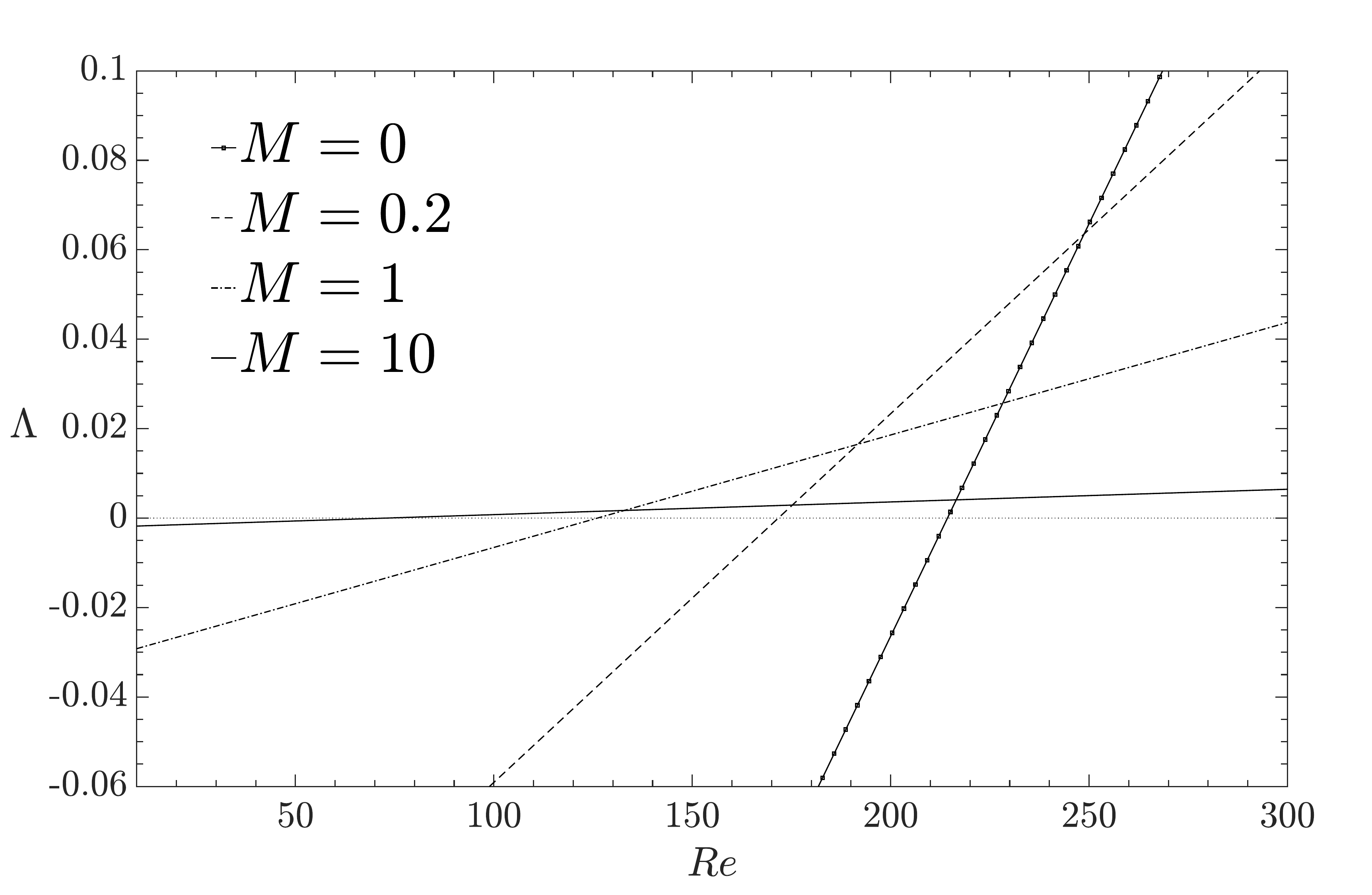}
\caption{Asymptotic solution for the growth rate $\Lambda$ against Reynolds number $Re$, plotted for $z_1=0.1$, $z_2 = 0.9$,  $\tilde{F}=1$ and varied $M$. }\label{growth_rate_plot}
\end{figure}

In figure \ref{growth_rate_plot},  we plot the growth rate $\Lambda$ against the Reynolds number for varied $M$.  Similar to our analysis of the critical Reynolds number,  our results are in good agreement with \citet{walters2018effect},  demonstrating that retaining higher-order azimuthal modes does not have a significant impact on the value of the growth rate.  

\section{Conclusions}\label{Conclusion_section}
In this paper, we have derived series representations for the oscillatory normal modes that describe the leading-order fluid--structure interaction of an initially elliptical thin-walled elastic tube conveying an incompressible viscous fluid.  The global energy budget was then used to determine the rate of growth or decay of these modes over a longer time scale.  Valid when the oscillations in the tube wall are of small amplitude and high frequency, our solutions enable the first formal analysis of the errors incurred by writing the solution as a function of only the first azimuthal eigenmode. 

In order to obtain the solutions, we coupled two existing fluid and solid mechanics models.  For the solid mechanics,  we used the model of \citet{Netherwood2023deformations},  which describes the pressure-induced wall deformations of an elastic tube in terms of a set of azimuthal eigenmodes.  For the fluid mechanics,  we used the long-wavelength high-frequency asymptotic model of \citet{whittaker2010predicting}.  The problem was then decomposed into steady and oscillatory parts. 

The oscillatory problem was found to admit normal-mode solutions
each containing a single frequency component. The problem for
each normal mode was formulated in terms of a set of pressure
modes, with the dimensionless oscillatory pressure decomposed as
$\hat{p} = \e^{i \omega t} \sum_{n=1}^{\infty} \tilde{p}_n(z)$.
Each pressure mode $\tilde{p}_n$ corresponds to the pressure
driving the axial sloshing flow due to the $n$th azimuthal solid
deformation eigenmode. We found that the azimuthal modes do not decouple at leading order. The governing equation for the component of the pressure associated with each individual eigenmode was forced by a pressure distribution made up of the sum of each of the modes,  and was shown to be of the form
\begin{equation}\label{nthmode_equation_coupled_conclusion}
\mathscr{L}(\tilde{p}_n)  = q_nt_n \sum_{i=1}^\infty \tilde{p}_i.
\end{equation}
To overcome difficulties induced by the coupling,  we exploited the fact that the product $q_nt_n$,  which multiplies the forcing in (\ref{nthmode_equation_coupled_conclusion}) decays rapidly with an increase in mode number, $n$. This enabled us to adopt a series expansion for the pressure and oscillation frequency.

With respect to a parameter,  $\epsilon$,  which is the limiting decay rate of $q_nt_n$ as $n \to \infty$,  the series representations for the oscillation frequency and pressure take the respective forms:
\begin{align}
\tilde{p}(z)& =p_{10}+\epsilon\left(p_{11}+p_{20} \right) + \epsilon^2(p_{12} + p_{21} + p_{30})+O(\epsilon^3), \label{full_p_conclusion}\\
\omega &=\omega_0+\epsilon \omega_1+ \epsilon^2 \omega_2 + O(\epsilon^3). \label{full_omega_conclusion}
\end{align}
Here $\omega_0$ is the leading-order oscillation frequency, with first correction $\omega_1$; $p_{10}$ is the leading-order component of the pressure associated with the first azimuthal mode, with correction $p_{11}$; and $p_{20}$ is the leading component of the pressure associated with the second azimuthal mode,  etc.  Our results demonstrate that errors associated with truncating (\ref{full_p_conclusion})--(\ref{full_omega_conclusion}) after $O(\epsilon^0)$ are typically small, and hence throughout most of $(M, \tilde{F}, \sigma_0)$ parameter space, the system is well approximated by $p_{10}$ and $\omega_0$. For certain limiting cases ($\tilde{F} \gg 1$ and $ M \ll 1)$, we found that errors associated with the oscillation frequency can grow to be as large as $O(1)$. However,  this parameter regime is not particularly relevant for the modelling of self-excited oscillations based on physical grounds,   since large $\tilde{F}$ corresponds to a situation in which azimuthal bending effects are small compared with axial curvature-tension effects,  and small $M$ represents negligible wall inertia.  

The normal-mode solutions described above are only the
leading-order solutions in an expansion in powers of dimensionless
amplitude $\delta$, aspect ratio $\ell^{-1}$, inverse Strouhal
number $St^{-1}$, and inverse Womersley number $\alpha^{-1}$.
The growth and decay of these modes over timescales larger than
the oscillation timescale would usually be determined by higher-order
corrections.  However,  rather than having to compute these corrections
explicitly, we were able to compute the growth rates in \S5 using
the global energy budget. This also allowed us to obtain a
stability criterion in the form of a critical Reynolds number
$\Rey_c$ for the mean flow. For $Re \leq Re_c$ the steady
system is stable to these normal-mode perturbations.  For $Re >
Re_c$, it is unstable to at least the fundamental normal mode.  Our results indicated that the higher-order azimuthal modes provide a negligible contribution to the critical Reynolds number,  which formally justifies the adhoc assumption invoked by \citet{rationalderivationofatubelaw} and \citet{walters2018effect}.

The long-wavelength high-frequency oscillations assumed for this model result in the tube's transmural pressure (at leading--order) being cross-sectionally uniform (i.e $\tilde{p} = \tilde{p}(z)$). A reasonable extension to the work presented in this article would be to include higher-order effects from the fluid mechanics, in which the pressure varies within the cross section. It would be simple to include such dependence when modelling the wall motion, since the results from \cite{Netherwood2023deformations} permit azimuthal variation in the transmural pressure.  Another way in which one could build upon the current work is by considering tubes with different initial cross-sectional shapes.  \citet{Netherwood2024thesis} showed that there exists a family of cross-sectional shapes for which an azimuthally uniform
transmural pressure forces only a single azimuthal eigenmode.
With such a cross-sectional shape, deformations with that
eigenmode could exist in the fluid-structure interaction problem
considered here without exciting any other modes, removing the
need for a series solution for the normal modes. Further details
are expected to be published shortly (\citet{Netherwood2025monomode}).  Other extensions and improvements include considering tubes with an initially axially non-uniform cross-sectional shape,  and a modification of the model to permit pressure--pressure boundary conditions.  
\section*{Acknowledgements}
DJN would like to acknowledge the financial support of the University of East Anglia to undertake the Ph.D. project of which this work is a part. 
\section*{Author contributions}
\begin{itemize}
\item \textbf{DJN -} Formal analysis,  Investigation, Methodology, Writing - original draft,  Writing - review and editing, software,  visualisation,  Data curation. 
\item \textbf{RJW-} Conceptualisation,  Investigation,  Methodology,  Project administration,  Writing - review and editing,  Supervision.  
\end{itemize}
\bibliographystyle{apalike}
\bibliography{Netherwood2025Starling}

\begin{appendix}
\section{Large-$n$ asymptotic behaviour of $q_nt_n$}\label{asymptotic_appendix}
The numerical results obtained by \cite{Netherwood2023deformations} show that the numerical constants
$q_n t_n$ decay rapidly as $n$ increases. In this Appendix,  we consider
their asymptotic behaviour as $n \to \infty$. We were unable to
determine the behaviour analytically,  so instead resort to fitting
functions to the numerical data. As we shall see below, the data
suggests an asymptotic relationship of the form
\begin{equation} q_n t_n \sim Q \epsilon^{n-1}
  \Bigl( 1 + cd^{-(n-1)} + \ldots \Bigr)
\end{equation}
where the constants $Q$, $\epsilon$, $c$ and $d$ are estimated
for different values of the ellipticity $\sigma_0$. (The decision to
use $n-1$, rather than $n$,  in the exponents is made for convenience
in the calculations in \S\ref{unsteady_coupled_problem_section}.)

\begin{figure}
\centering 
\includegraphics[width = \textwidth]{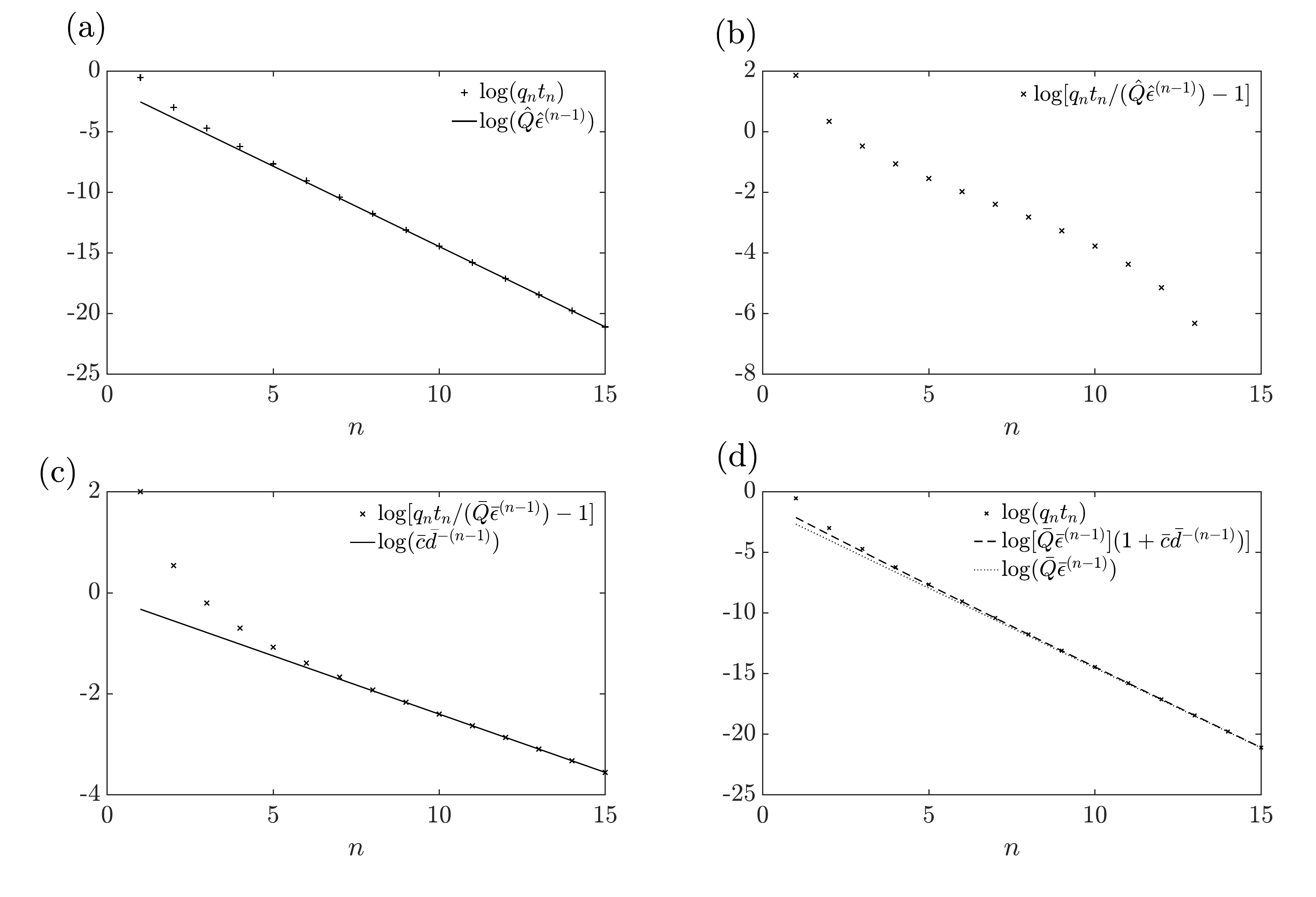}
\caption{The stages of the fitting process examining the large-$n$ asymptotic behaviour of $q_nt_n$ for the case $\sigma_0 = s_2 = 0.6$. The estimates $\hat{Q},  \hat{\epsilon},  \bar{Q},  \bar{\epsilon},  \bar{c},  \bar{d}$ have been found by fitting appropriate functions to the final points in the data.  Details of the specific fitting process are described in Appendix \ref{asymptotic_appendix}.  \label{fig:q_nt_n_s2_asymptotics} }
\end{figure}

Figure \ref{fig:q_nt_n_s2_asymptotics}(a) shows a plot of $\log(q_n t_n)$
against $n$. The approximately linear behaviour for large $n$ suggests
an asymptotic relationship of the form $q_n t_n = Q \epsilon^{n-1}$.
The straight line shown is a fit of this function to the final two
data points $n=14,15$. This gives the estimates $\hat{Q}$ and
$\hat{\epsilon}$.

Figure \ref{fig:q_nt_n_s2_asymptotics}(b) shows a plot of the log of the
relative error $q_n t_n/(\hat{Q}\hat{\epsilon}^{n-1})-1$ from this
fit against $n$. We observe approximately linear behaviour for
intermediate $n$. This suggests a relationship of the form
$q_n t_n/Q\epsilon^{n-1}-1 \sim cd^{-(n-1)}$. (With this relationship,
we would not expect linear behaviour over the whole of the domain,
since for smaller $n$, $n$ is not large; while for larger $n$, the errors
arising from the errors in the estimate $\hat{Q}$ and $\hat{\epsilon}$
will dominate.) 

Fitting $q_n t_n = Q\epsilon^{n-1} ( 1 +
cd^{-(n-1)})$ to the final four data points $n=12,13,14,15$ yields estimates $\bar{Q}$, $\bar{\epsilon}$, $\bar{c}$, $\bar{d}$. In
figure \ref{fig:q_nt_n_s2_asymptotics}(c),  we show the relative error from
the leading-order term $\bar{Q}\bar{\epsilon}^{n-1}$ and the curve
from the full fitted function, to demonstrate how this function does
indeed capture the error as $n \to \infty$.

Finally, in figure \ref{fig:q_nt_n_s2_asymptotics}(d),  we show the fitted
functions on the original logarithmic axes. The estimates for
$Q,\epsilon,c,d$ that were calculated for the representative values of
$\sigma_0$ are shown in Table \ref{Q_epsilon_table_values}.

\begin{table}
\centering
\begin{tabular}{P{0.5in}|P{0.5in}lP{0.5in}lP{0.5in}lP{0.5in}l}
\hline
           $\sigma_0$  & $~s_1$ & $~~~s_2$ & $s_3$ & $ ~~~s_4$ \\ 
           &0.9540 & 0.6000& 0.3840& 0.2194 \\\hline
$\hat{Q}$        & 0.0345           & 0.0782           & 0.0843          & 0.0734           \\ 
$\hat{\epsilon}$ & 0.0704          & 0.2657           & 0.5275           & 0.7581           \\ 
$\bar{Q}$        & 0.0313           & 0.0688          & 0.0715           & 0.0366           \\ 
$\bar{\epsilon}$ & 0.0711           & 0.2676           & 0.5325           & 0.7889          \\ \hline
\end{tabular}
\caption[Estimates of the numerical parameters $Q$ and $\epsilon$.]{Estimates of the numerical parameters $Q$ and $\epsilon$ for the representative values of $\sigma_0 = s_i$.  } \label{Q_epsilon_table_values}
\end{table}
\section{Oscillatory energy in the tube wall}\label{sec:energy_appendix}
In this appendix we adapt the methodology set out in \cite{whittaker2010predicting} and \cite{walters2018effect} to determine an expression for $\tilde{\mathbb{E}}_s$,  the total period-averaged energy (elastic and kinetic) in the wall as a result of all of the azimuthal modes.

Following \cite{walters2018effect} and \cite{whittaker2010predicting}, the dimensionless rate of working on the tube wall by the transmural pressure is
\begin{equation}\label{dimensionless_rate_of_working_Es}
\diff{E_s}{t} = \Delta St^2 \ell^2  \int_{z_1}^{z_2} p_{\mathrm{tm}} \pdiff{A}{t} \D{z}. 
\end{equation}
Nondimensionalising (\ref{A^star_series_expression}) and substituting into (\ref{dimensionless_rate_of_working_Es}) gives
\begin{equation}\label{dimensionless_rate_of_working_Es11}
\diff{E_s}{t} = \Delta St^2 \ell^2  \int_{z_1}^{z_2} p_{\mathrm{tm}} \sum_{n=1}^\infty \pdiff{A_n}{t}\D{z} = \sum_{n=1}^\infty \Delta St^2 \ell^2  \int_{z_1}^{z_2} p_{\mathrm{tm}}\pdiff{A_n}{t}\D{z}.  
\end{equation}

From (\ref{tubelaw_coupled}),  we can obtain an explicit uncoupled relationship between each of the area change components $A_n$ and the global dimensionless transmural pressure.  Using the dimensionless form of (\ref{tubelaw_coupled}) to eliminate $p_{\mathrm{tm}}$ from (\ref{dimensionless_rate_of_working_Es11}),  we obtain
\begin{equation}\label{E_s_2}
\diff{E_{s}}{t} = \sum_{n=1}^\infty \frac{St^2 \ell^2}{q_nt_nA_0} \int_{z_1}^{z_2}  \left( \lambda_n A_n +M\pdiff[2]{A_n}{t} - \tilde{F}\pdiff[2]{A_n}{z}\right) \pdiff{A_n}{t}\D{z}.
\end{equation}

Each term in the integral can be written as a total derivative,  so we can integrate (\ref{E_s_2}) once in time to yield
\begin{equation}\label{E_s_3}
E_{s} =\sum_{n=1}^\infty \frac{St^2 \ell^2}{2 q_nt_nA_0}\int_{z_1}^{z_2}\lambda_n A_n^2 +M\left(\pdiff{A_n}{t}\right)^2+\tilde{F}\left(\pdiff{A_n}{z} \right)^2 \D{z} + C,
\end{equation}
where $C$ is an arbitrary constant.

Recalling expression (\ref{area_expression_A_n}),  have that 
\begin{equation}\label{A_n_appendix_dimensionless}
A_n(z,t) = \frac{1}{\alpha^2 \ell St} \bar{A}_n(z) + \Delta(t) \hat{A}_n(z,t).
\end{equation}
Writing $\hat{A}_n = Re(\tilde{A}_n(z)\exp{(i\omega t)})$ in (\ref{A_n_appendix_dimensionless}) and substituting into (\ref{E_s_3}) gives
\begin{align}
E_{s} & =\sum_{n=1}^\infty \frac{St^2 \ell^2}{2 q_nt_nA_0}\int_{z_1}^{z_2}\lambda_n \left[\frac{1}{\alpha^2 \ell St} \bar{A}_n(z) + \Delta(t) Re(\tilde{A}_n(z)e^{i\omega t}) \right]^2 \nonumber \\ 
& \qquad  +M\left[\Delta(t) Re(\tilde{A}_n(z)i \omega e^{i\omega t}) \right]^2 \nonumber \\
& \qquad \qquad +\tilde{F}\left[\frac{1}{\alpha^2 \ell St} \diff{\bar{A}_n}{z} + \Delta(t) Re\left(\diff{\tilde{A}_n}{z}e^{i\omega t}\right) \right]^2 \D{z} + C. \label{E_s_4}
\end{align}
From \cite{walters2018effect} and \cite{whittaker2010predicting},  we average $E_s$ over the timescale of a single oscillation as follows
\begin{equation}\label{avergae_single_oscillation_operator}
\mathbb{E}_s = \langle E_s \rangle = \frac{\omega}{2 \pi} \int_0^{\frac{2\pi}{\omega}} E_s \D{t}.
\end{equation}
We note also that a function decomposed as $\mathbb{A}(z,t) = \bar{\mathbb{A}}(z) + Re(\tilde{\mathbb{A}}(z)\exp{(i \omega t)})$ satisfies
\begin{equation}\label{square_average_identity}
\langle \mathbb{A}^2\rangle = \bar{\mathbb{A}}^2+\frac{1}{2}|\tilde{\mathbb{A}}|^2. 
\end{equation}
We now take the average of (\ref{E_s_4}) and apply (\ref{square_average_identity}) to obtain
\begin{align}\label{E_s_5}
\mathbb{E}_{s} & = \sum_{n=1}^\infty\frac{St^2 \ell^2}{2 q_n t_n A_0} \int_{z_1}^{z_2} \lambda_n \left(\frac{1}{\alpha^4 \ell^2 St^2}\bar{A}_n^2+\frac{\Delta^2}{2}|\tilde{A_n}|^2 \right) + \frac{M \Delta^2 \omega^2 }{2}|\tilde{A}_n|^2 \nonumber \\
& \qquad \qquad + \tilde{F}\left(\frac{1}{\alpha^4 \ell^2 St^2}\left(\diff{\bar{A}_n}{z} \right)^2 +\frac{\Delta^2}{2}\left(\diff{\tilde{A}_n}{z} \right)^2\right)\D{z} + C.
\end{align}
Retaining only oscillatory contributions in (\ref{E_s_5}),  we find that the dimensionless mean energy in the tube wall due to oscillations is given by
\begin{align}\label{E_s_6}
\tilde{\mathbb{E}}_{s} & =\sum_{n=1}^\infty \frac{\Delta^2 St^2 \ell^2}{4 q_n t_n A_0} \int_{z_1}^{z_2} \left(\lambda_n + M \omega^2  \right)|\tilde{A}_n|^2 + \tilde{F}\bigg| \diff{\tilde{A}_n}{z} \bigg|^2\D{z}.
\end{align}
Since the expressions present in (\ref{energy_budget})--(\ref{Dissipation_energy}) are written in terms of the pressure,  it is convenient to write (\ref{E_s_6}) in terms of the pressure also.  To do this,  we re-introduce the pressure modes $\hat{p}_n = Re(\tilde{p}_n(z)\exp{(i\omega t)})$ from \S\ref{flexible_equations_section},  which can be related to the area-change modes $\hat{A}_n = Re(\tilde{A}_n\exp{(i\omega t)})$ using (\ref{coupling_p_n_hat}) as follows
\begin{equation}\label{Anpn_coupling appendix}
\tilde{A}_n = -\frac{A_0 \tilde{p}_n''}{\omega^2}.
\end{equation}
Substituting (\ref{Anpn_coupling appendix}) into (\ref{E_s_6}) yields
\begin{equation}\label{E_s_7}
\tilde{\mathbb{E}}_{s} = \sum_{n=1}^\infty\frac{\Delta^2 St^2 \ell^2A_0}{4 q_nt_n \omega^4}\int_{z_1}^{z_2} \left(\lambda_n +M\omega^2 \right)\tilde{p}''_n(\tilde{p}''_n)^\dagger+\tilde{F}\tilde{p}'''_n(\tilde{p}'''_n)^\dagger \D{z},
\end{equation} 
where the superscript $\dagger$ represents the complex conjugate. 

By adapting the methodology set out in \cite{walters2018effect},  we can write the integral in (\ref{E_s_7}) in a more convenient form.  Integrating the second term in (\ref{E_s_7}) by parts and noting that $\tilde{p}_n''=0$ at $z = z_1,z_2$,  we have that
\begin{equation}\label{E_s_8}
\tilde{\mathbb{E}}_{s} = \sum_{n=1}^\infty\frac{\Delta^2 St^2 \ell^2A_0}{4 q_nt_n \omega^4}\int_{z_1}^{z_2} \left(\lambda_n +M\omega^2 \right)\tilde{p}''_n(\tilde{p}''_n)^\dagger-\tilde{F}\tilde{p}''''_n(\tilde{p}''_n)^\dagger \D{z}.
\end{equation} 
Writing (\ref{flexible_governing}) in terms of $z$ and using (\ref{pressure_expansion}) to write the forcing in terms of the global axial mode of the oscillatory fluid pressure $\tilde{p}$,  it follows that
\begin{equation}\label{expr_for_p''''}
-\tilde{F}\tilde{p}''''_n = (M\omega^2-\lambda_n)\tilde{p}''_n-q_nt_n \omega^2 \tilde{p}.
\end{equation}
On multiplying both sides of (\ref{expr_for_p''''}) by $(\tilde{p}_n'')^\dagger$ and substituting into (\ref{E_s_8}),  we obtain
\begin{equation}\label{E_s_9}
\tilde{\mathbb{E}}_{s} = \sum_{n=1}^\infty\frac{\Delta^2 St^2 \ell^2A_0}{4 \omega^2}\int_{z_1}^{z_2} \frac{2M}{q_nt_n}\tilde{p}_n''(\tilde{p}_n'')^\dagger-\tilde{p}(\tilde{p}_n'')^\dagger \D{z}.
\end{equation}
Distributing the sum in (\ref{E_s_9}) and noting that $\sum_{n=1}^\infty (\tilde{p}_n'')^\dagger = (\tilde{p}'')^\dagger$,  we find that
\begin{equation}\label{E_s_10}
\tilde{\mathbb{E}}_{s} = \frac{\Delta^2 St^2 \ell^2A_0}{4 \omega^2}\int_{z_1}^{z_2}\sum_{n=1}^\infty \frac{2M}{q_nt_n}\tilde{p}_n''(\tilde{p}_n'')^\dagger-\tilde{p}(\tilde{p}'')^\dagger \D{z}.
\end{equation}
Since $\tilde{p}''=\tilde{p}''_n=0$ in the rigid sections,  we extend the domain of integration in (\ref{E_s_10}) to the interval $z \in (0,1)$ and then integrate the second term in the integral by parts and apply the endpoint boundary conditions to obtain
\begin{equation}\label{E_s_12}
\tilde{\mathbb{E}}_{s} = \frac{\Delta^2 St^2 \ell^2A_0}{4 \omega^2}\int_{0}^{1}|\tilde{p}'|^2 + \sum_{n=1}^\infty \frac{2M}{q_nt_n}\tilde{p}_n''(\tilde{p}_n'')^\dagger\D{z}.
\end{equation}
Finally,  on writing $\tilde{p}' = \sum_{n=1}^\infty \tilde{p}_n'$,  we obtain
\begin{equation}\label{E_s_13}
\tilde{\mathbb{E}}_{s} = \frac{\Delta^2 St^2 \ell^2A_0}{4 \omega^2}\int_{0}^{1} \bigg|\sum_{n=1}^\infty\tilde{p}_n'\bigg|^2 + \sum_{n=1}^\infty \frac{2M}{q_nt_n} |\tilde{p}_n''|^2\D{z}.
\end{equation}
\end{appendix}
\end{document}